\documentclass{mn2e}

\usepackage[dvips]{graphicx}

\newcommand{\asm}{{\it RXTE}/ASM }
\newcommand{\vela}{{\it Vela 5B}/ASM }

\newcommand{\batse}{{\it CGRO}/BATSE }
\newcommand{\ariel}{{\it Ariel 5}/ASM }
\newcommand{\ariell}{{\it Ariel 5}/ASM}
\newcommand{\asmm}{{\it RXTE}/ASM}
\newcommand{\ginga}{{\it Ginga}/ASM }
\newcommand{\gingaa}{{\it Ginga}/ASM}
\newcommand{\rxte}{{\it RXTE} }

\newcommand{\dayys}{d$^{-1}$}
\newcommand{\be}{\begin{equation}}
\newcommand{\ee}{\end{equation}}
\newcommand{\msun}{{{\rm M}_{\sun}}}

\title[Periodic long-term variability of Cygnus~X-1]
      {Periodic long-term X-ray and radio variability of \\ Cygnus~X-1}

\author
 [Lachowicz et al.]
{\noindent
 Pawe{\l} Lachowicz,$^1$\thanks{E-mail: (paulo, aaz, alex)@camk.edu.pl}
 Andrzej A.~Zdziarski,$^1$\footnotemark[1]
 Alex Schwarzenberg-Czerny,$^{1,2}$\footnotemark[1]
 \newauthor
 Guy G. Pooley$^3$ and
 Shunji Kitamoto$^4$ \\
 $^1$Centrum Astronomiczne im.\ M. Kopernika, Bartycka 18, 00-716 Warszawa, Poland\\
 $^2$Obserwatorium Astronomiczne, Uniwersytet A. Mickiewicza, S{\l}oneczna 36,
     60-286 Pozna\'n, Poland \\
 $^3$Cavendish Laboratory, J. J. Thomson Avenue, Cambridge CB3 0HE \\
 $^4$Department of Physics, Rikkyo University, 3-34-1 Nishi-Ikebukro,
     Toshima-ku 171-8520, Japan \\
}

\date{Accepted 2006 July 21. Received 2006 February 14; in original form 2005 August 15}

\pagerange{\pageref{firstpage}--16}
\pubyear{2006}

\begin{document}

\maketitle

\label{firstpage}

\begin{abstract}
We present a comprehensive analysis of long-term periodic variability of Cyg X-1 using the method of multiharmonic analysis of variance applied to available monitoring data since 1969, in X-rays from {\it Vela 5B, Ariel 5, Ginga, CGRO\/} and {\it RXTE\/} satellites and in radio from the Ryle and Green Bank telescopes. We confirm a number of previously obtained results, and, for the first time, find an orbital modulation at 15 GHz in the soft state and show the detailed non-sinusoidal shape of that modulation in the hard state of both the 15-GHz emission and the X-rays from the {\it RXTE}/ASM. We find the {\it CGRO}/BATSE data are consistent with the presence of a weak orbital modulation, in agreement with its theoretical modelling as due to Compton scattering in the companion wind. We then confirm the presence of a $\sim$150-d superorbital period in all of the data since $\sim$1976, finding it in particular for the first time in the {\it Ariel 5\/} data. Those data sets, covering $>$65 superorbital cycles, show a remarkable constancy of both the period and the phase. On the other hand, we confirm the presence of a $\sim$290-d periodicity in the 1969--1979 {\it Vela 5B\/} data, indicating a switch from that period to its first harmonic at some time $\la$1980. We find the superorbital modulation is compatible with accretion disc precession. Finally, we find a significant modulation in the {\it RXTE}/ASM data at a period of 5.82~d, which corresponds to the beat between the orbital and superorbital modulations provided the latter is prograde.
\end{abstract}


\section{Introduction}
\label{s:intro}

Cyg X-1, a persistent Galactic black-hole binary, was discovered in 1964 June
(Bowyer et al.\ 1965), and it has been extensively studied since then. Its
orbital period is $P_{\rm orb}=5.6$~d and the optical component, HDE
226868/V1357 Cygni (Bolton 1972; Webster \& Murdin 1972) is an OB
supergiant (Walborn 1973). The masses of the two stars remain the subject of some dispute, with the most recent determination of $40\pm 10\msun$ and $20\pm
5\msun$ for the supergiant and the black hole, respectively (Zi\'o{\l}kowski
2005).

Focused wind is the main accretion channel, though the companion nearly
fills its Roche lobe (Gies \& Bolton 1986; Gies et al.\ 2003, hereafter G03).
Bound-free absorption by the wind leads to a strong modulation with the orbital period of the X-rays (Priedhorsky, Brandt \& Lund 1995; Zhang, Robinson \& Cui 1996; Wen et al.\ 1999, hereafter W99; Brocksopp et al.\ 1999a, hereafter B99; Kitamoto et al.\ 2000, hereafter K00) although the modulation is relatively irregular, manifesting itself by dips in the X-ray light curve near the superior conjunction of the black hole (Ba{\l}uci\'nska-Church et al.\ 2000). The modulation is also seen in the radio (Pooley, Fender \& Brocksopp 1999, hereafter P99; B99), being then due to free-free absorption by the wind (Brocksopp, Fender \& Pooley 2002). The optical emission is also modulated with the orbital period, showing two minima/maxima per period due to the ellipsoidal shape of the star (Lyutyi 1985; Kemp 1987; Voloshina, Lyutyi \& Tarasov 1997; Brocksopp et al.\ 1999b).

Typical of X-ray binaries, Cyg X-1 shows two main spectral states, hard and soft
(e.g., Gierli{\'n}ski et al.\ 1997, 1999; Frontera et al.\ 2001; McConnell et al.\ 2002; Zdziarski \& Gierli{\'n}ski 2004). Interestingly, the orbital modulation disappears in the soft state (W99), probably due to strong photoionization of the wind by the enhanced soft X-ray flux.

Cyg X-1 was also found to display superorbital modulation, i.e., at much longer
time scales than $P_{\rm orb}$. Originally, a period of 294 d was reported by
Priedhorsky, Terrell \& Holt (1983, hereafter PTH83) in X-rays and by Kemp et
al.\ (1983) in the optical. Later, however, a $\sim$150 d period from the radio
to X-rays was found by numerous authors (B99; P99; K00; {\"O}zdemir \& Demircan 2001; Karitskaya et al.\ 2001; Benlloch et al.\ 2001, 2004). The probable cause of this modulation is precession of the accretion disc.

We present here a comprehensive analysis of the orbital and superorbital modulation of Cyg X-1 in the X-ray and radio bands. We use most of
the available X-ray and radio monitoring data, applying to it the method of multiharmonic analysis of variance (hereafter abbreviated as mhAoV). Section \ref{s:selection} presents details of our data selection, and Section \ref{s:ta}, our analysis method. Sections \ref{s:orb} and \ref{superorbital} present our results on the orbital and superorbital modulation, respectively. Discussion and theoretical interpretation are given in Section \ref{s:discussion}, and our conclusions, in Section \ref{s:conclusions}. In Appendix A, we separately reanalise the presence of the $\sim$290-d periodicity in the {\it Vela 5B\/} data. In Appendix B, we use the method of structure function as an independent check of our results. 

\section{Data Selection}
\label{s:selection}

\begin{figure*}
\begin{center}
\vspace*{5pt}
\includegraphics[height=14.5cm,angle=-90]{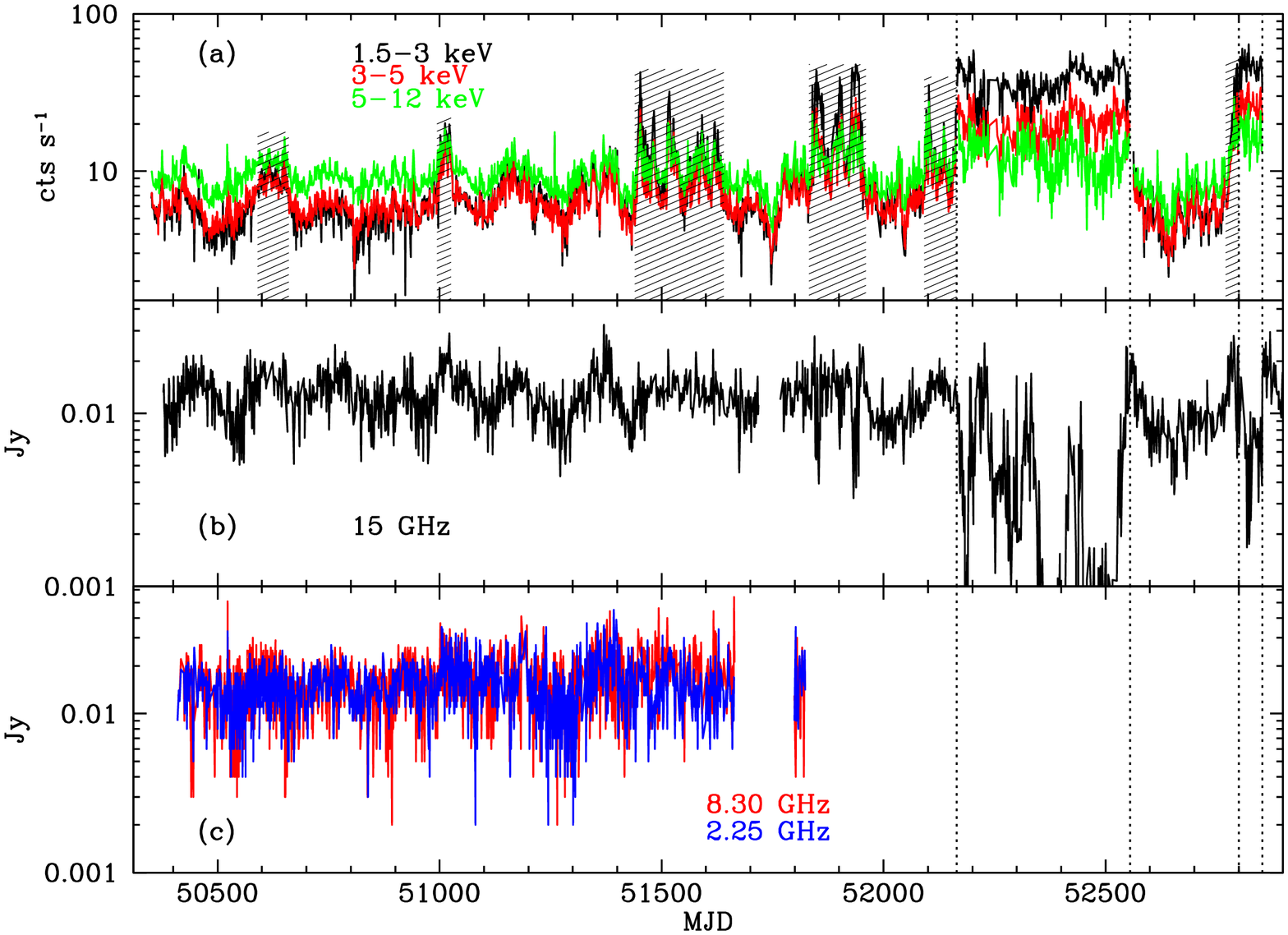}
\vspace*{0pt}
{
\includegraphics[height=14.8cm,angle=-90]{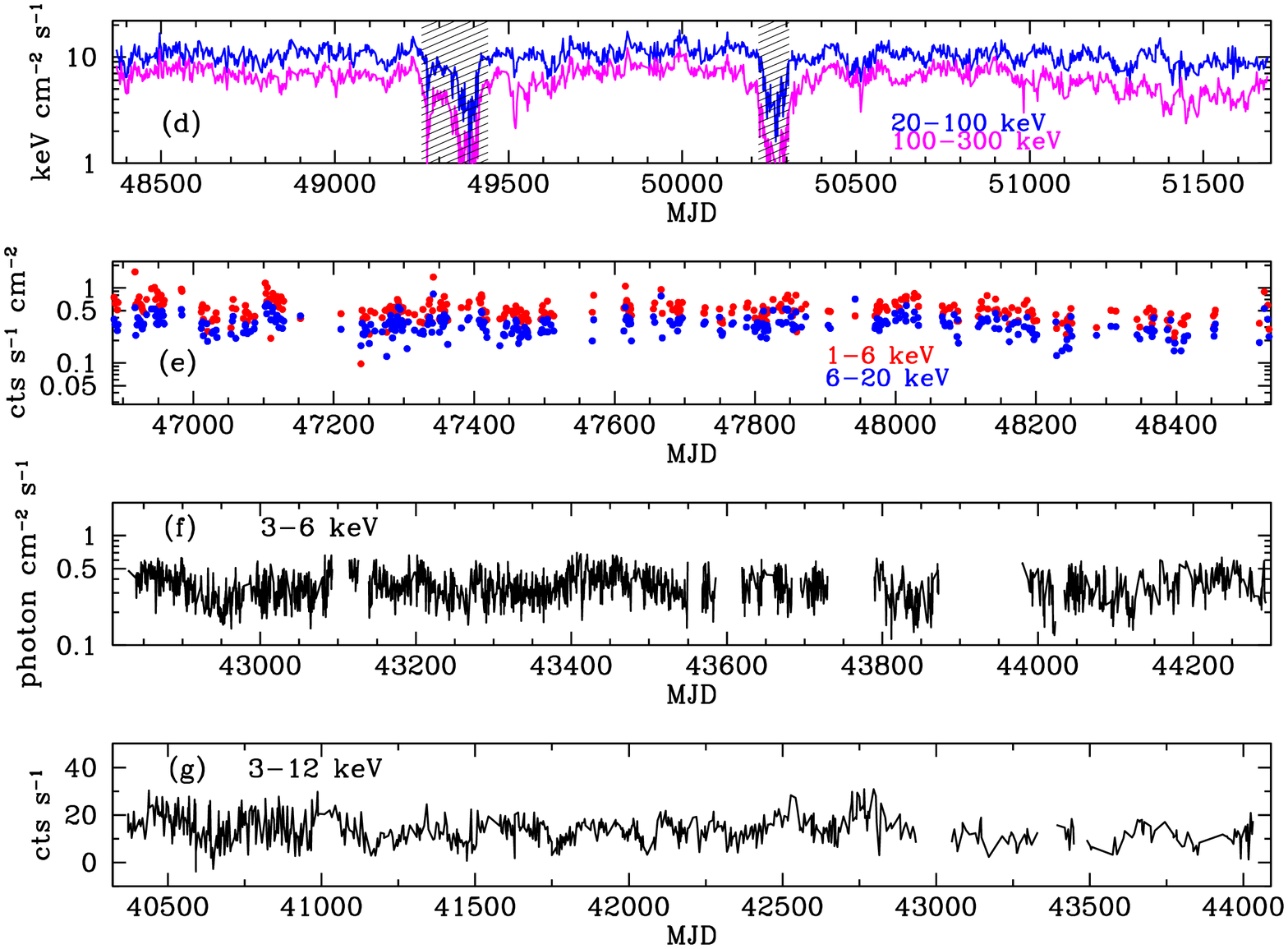}
}
\end{center}
\caption{
{\it (a)} The \asm 1-day average light curves corresponding to hard states and the soft states of 2001--2002 and 2003. The used boundaries of the soft states
are marked by dotted lines. The shaded regions correspond to the data not taken
into account in our analysis in order to provide approximately the same flux
level. {\it (b)} The 1-day average 15 GHz light curve from the Ryle telescope. {\it (c)} The 2.25 and 8.30 GHz GBI data. {\it (d)} The \batse light curves. The shaded regions correspond to the 1994 and 1996 soft states, which are not included in the analysis. {\it (e)} The \ginga data. {\it (f)} The \ariel light curve. {\it (g)} The \vela 0.25-day average light curve.
}
\label{f:lc}
\end{figure*}

We analyse available light curves separately for the hard and soft states. We follow the definitions of those states of Zdziarski et al.\ (2002); thus the state boundaries may slightly change if other definitions are used. We use X-ray data  from the Burst and Transient Source Experiment aboard {\it Compton Gamma Ray  Observatory\/} ({\it CGRO}/BATSE; Paciesas et al.\ 1997), and the All-Sky Monitors (ASM) aboard {\it Rossi X-ray Timing Explorer\/} ({\it RXTE}; Bradt, Rothschild \& Swank 1993; Levine et al.\ 1996), {\it Ginga\/} (Makino et al.\ 1987; Tsunemi et al.\ 1989), {\it Ariel 5\/} (Holt 1976), and {\it Vela-5B\/} (Conner, Evans \& Belian 1969). We use the 15-GHz radio data from the Ryle Telescope of the Mullard Radio Astronomy Observatory, and the 2.25 and 8.30 GHz data from the Green Bank Interferometer (GBI) of the National Radio Astronomy Observatory in Green Bank, WV. Table \ref{t:lc} gives the log of the data, which are presented graphically in Fig.\ \ref{f:lc}.

\begin{table}
\caption{The log of the light curves used in this work. H and S refer to the hard and soft spectral state, respectively. The two sub-intervals of the hard state from the {\it RXTE}/ASM and Ryle instruments are treated jointly in the analysis. See Section \ref{s:selection} for the criteria of data selection and for some additionally rejected intervals.
 }
\begin{center}
\begin{tabular}{clccccc}
\hline
State & Instrument  & Start  & End  & Time span \\
&             & (MJD) & (MJD) & (d) &   \\
\hline
H      & \asm   & 50350 & 52099  & 1750 \\
       &        & 52565 & 52770  & 206  \\
       & Ryle   & 50377 & 52164  & 1788 \\
       &        & 52565 & 52770  & 206  \\
       & GBI    & 50409 & 51823  & 1415 \\
       & \batse & 48371 & 51686  & 3316  \\
       & \ginga & 46887 & 48531  & 1645  \\
       & \ariel & 42830 & 44292  & 1463  \\
       & \vela  & 40368 & 44042  & 3675  \\
\hline
S      & \asmm, Ryle  & 52165 & 52555 & 391 \\
       &              & 52800 & 52853 &  54 \\
\hline
\label{t:lc}
\end{tabular}
\vspace*{-15pt}
\end{center}
\end{table}

For the {\it RXTE}/ASM, we use the dwell count rates from the Definitive Products database of the High Energy Astrophysics Science Archive Research Center (HEASARC). Each dwell lasts $\sim$90 s (Levine et al.\ 1996) and its number within one day varies up to 30 for Cyg X-1. For the hard state, we have excluded high-flux intervals dominated by the so-called failed state transitions (Pottschmidt et al.\ 2003). In this selection procedure, we have chosen MJD 50660--50990 as the reference interval, during which the count rate in each of the three energy bands (1.5--3, 3--5 and 5--12 keV) remained within $4\sigma$ around the respective mean. Then, we have searched 30-day intervals in the remaining parts of the hard state, and have excluded the intervals with more than 40 per cent of points in that interval exceeding the $4\sigma$ level. This, with some further minor adjustments, resulted in the intervals of MJD 50590--50660, 50995--51025, 51440--51640 and 51840--51960 excluded from the first part of the hard state of Table \ref{t:lc}. We have furthermore omitted negative count rates, which comprise $<$0.04 of the ASM data points, and thus their omission has a negligible effect.  We have also applied the barycenter correction to the ASM times of measurements; however, its effect is negligible. We note that the behaviour of Cyg X-1 after MJD 52853 until now ($\sim$53600) has not been suitable for our analysis as showing alternate periods of the hard and soft states of rather short durations.

We notice that there is no agreement in the literature on the definition of the soft state. For example, Belloni et al.\ (1996) called the 1996 soft state of Cyg X-1 `intermediate' whereas numerous other authors termed it `soft'. In particular, the two occurences of the soft state considered by us (see Table \ref{t:lc}), while characterized by the relatively high and steady high X-ray flux (see Fig.\ \ref{f:lc}a), may be classified as containing also the intermediate state based on some other criteria. In general, blackbody-disc dominated states of black-hole binaries range between ultrasoft, with almost no high-energy tail, to very high or intermediate, where the high-energy tail starts at the top of the disc blackbody (see, e.g., Zdziarski \& Gierli\'nski 2004 for a review), with different timing properties of the disc blackbody and the tail. The soft state of Cyg X-1 lies inbetween the ultrasoft and very high states, see, e.g., the spectra in Gierli\'nski et al.\ (1999), Frontera et al.\ (2001), Gierli\'nski \& Zdziarski (2003). 

We use the same BATSE data as Zdziarski et al.\ (2002). They contain 2729 days
of usable observations between 1991 April 25 and 2000 May 22 in the energy bands
of 20--100 keV and 100--300 keV. However, we exclude the periods of the soft state of 1994 and 1996, i.e., MJD 49250--49440 and MJD 50230--50307. Both of those occurences of the soft state are of relatively short duration. For the former, only the BATSE data are available, and the latter has already been extensively studied, with W99 finding no orbital modulation in the {\it RXTE}/ASM data.

We use the \ariel 3--6 keV and \vela 3--12 keV (averaged prior to analysis over 0.25-d intervals) light curves from the HEASARC database. We note that we use the {\it Ariel 5\/} data  (Table \ref{t:lc}) excluding the MJD 42338--42829, which were dominated by soft flare events and included the soft state of 1975 (Liang \& Nolan 1984). We have also omitted data points with negative fluxes. We then use the same \ginga 1--20 keV data as K00. They contain 339 measurements during 1987 February--1991 October.

The Ryle radio data used by us contain 10-min.\ average flux measurements. Parts of this data set have been studied by P99 and Benlloch et al.\ (2004). We have divided the data into the hard and soft states (Table \ref{t:lc}) based on the \asm data. Interestingly, we see a relavitely high level of the 15-GHz emission during the soft states. The 2.25 and 8.30 GHz GBI data (with the average sampling interval of $\sim$0.9 d; shown in Fig.\ \ref{f:lc} with $1\sigma$ errors) correspond to the hard state only. They have been screened according to the standard procedure\footnote{ftp://ftp.gb.nrao.edu/pub/fghigo/gbidata/gdata/00README}. An early part of the GBI data set has been studied by P99 and B99. 

\section{Time Analysis}
\label{s:ta}

In our analysis, we first determine the significance of the presence of a period in a light curve using the method of multiharmonic analysis of variance (mhAoV; Section \ref{ss:mdet}). We use the logarithm of either flux or count rate, $G(t)\equiv \ln {\cal F}(t)$, similarly to the use of magnitudes in optical astronomy. We do not take into account measurement errors in the analysis.

For a formulation and derivation of the quantitative results of the mhAoV method, see Schwarzenberg-Czerny (1999) and references therein. Then, for an established period, we fold and average the light curve and fit it with a Fourier series, which enables us to find the detailed shape of a periodic modulation including its phase (Section \ref{ss:fits}).

\subsection{The method of period detection}
\label{ss:mdet}

\subsubsection{General formulation and model orthogonality}
\label{ss:mdet:orth}

In general, we learn from experiments by fitting data, $x$, with a model, 
$x_{\|}$. The data contain $n$ measurements, and the model, $n_{\|}$ free 
parameters. The consistency of the data with the model is measured by a 
function, $\Theta$, called a statistic. A given model, $x_{\|}$, using a given 
statistic (e.g., $\chi^2$), yields its particular value, $\Theta_1$. Various 
methods used in the analysis of time series differ both in their choice of the 
model and the statistic; hence are difficult to compare directly. To enable such 
a comparison and for determining the significance of results,  $\Theta$ is 
converted into the false alarm probability, $P_1$. This is done considering a 
hypothetic situation, $H_1$, in which $x$ is pure white noise. Then each pair 
$(x_{\|},\Theta)$ corresponds to certain cumulative probability distribution of 
$\Theta$, namely $P(n_{\|},n;\Theta)$, with $P_1$ being the tail probability 
that under the hypothesis $H_1$ the experiment yields $\Theta>\Theta_1$, i.e., 
$P_1(\Theta>\Theta_1) = 1-P(n_{\|},n;\Theta_1)$.

Up to here, we have just outlined the classical Neyman-Pearson procedure of 
statistics. The specific method for analysis of time series used here differs 
from those commonly encountered in astronomy only in the choices of $x_{\|}$ and 
$\Theta$. Then, our accounting for variance, correlation and multiple 
frequencies in calculating $P$ is dictated by the laws of statistics. The 
probabilities derived by us from the data are the false alarm probabilities. 
However, we also call them below just probabilities or significance levels.

We note then that Fourier harmonics are not orthogonal in terms of the scalar 
product with weights at unevenly distributed observations. Certain statistical 
procedures employing classical probability distributions hold for orthogonal 
models only and fail in other cases. To avoid that, a popular variant of the 
power spectrum, Lomb (1976) and Scargle (1982, hereafter LS) periodogram relies 
on a special choice of phase such that the sine and cosine functions become 
orthogonal. We extend this approach by employing Szeg\"{o} orthogonal 
trigonometric polynomials as model functions. A series of $n_{\|}=2N+1$ 
polynomials corresponds to the orthogonal combinations of the $N$ lowest Fourier 
harmonics (Schwarzenberg-Czerny 1996). Orthogonal series are optimal from the 
statistical point of view because, by virtue of the Fisher lemma (Fisz 1963; 
Schwarzenberg-Czerny 1998), they guarantee the minimum variance of the fit 
residuals for a given model complexity (given by $n_{\|}$). Szeg\"{o} 
polynomials are also convenient in computations since the least-square solution 
may be obtained using recurrence and orthogonal projections, resulting in high 
computational efficiency, with the number of steps $\propto N$ instead of $N^3$ 
for $N$ harmonics.

\subsubsection{Variance, the AoV statistics, and model complexity}
\label{ss:mdet:var}

The LS method employs the sine as a model, and the quadratic norm, 
$\Theta_{\chi^2}=\|x-x_{\|}\|^2$, as the statistic. The corresponding 
probability distribution is $\chi^2$ with 2 degrees of freedom. Prior to use of 
the $\chi^2$ distribution, $\Theta_{\chi^2}$ has to be divided by the signal 
variance, $V$. However,$V$ is usually not known and has to be estimated from the 
data themselves. Then, neither $\Theta_{\chi^2}$ and variance estimates are 
independent nor their ratio follows the $\chi^2$ distribution, which effect has 
to be accounted for. A simple way to do it is to apply the Fisher Analysis of 
Variance (AoV) statistic, $\Theta\equiv (n-n_{\|}) \|x_{\|}\|^2/ (n_{\|}\|x - 
x_{\|}\|^2)$. Hence we call our method, involving Szeg\"{o} polynomials model 
and the AoV statistics, the multi-harmonic analysis of variance or mhAoV 
(Schwarzenberg-Czerny 1996). The probability distribution is then the 
Fisher-Snedecor distribution, $F$, rather then $\chi^2$, and $P_1= 
1-F(n_{\|},n_{\perp};\Theta)$ where $n_{\perp}=n-n_{\|}$. For everything else fixed, replacing $\chi^2$ with $F$ for $n=100$ yields an increase of $P_1(\chi^2)=0.001$ to $P_1(F)=0.01$. Thus, accounting for the unknown variance yields the mhAoV detection less significant, but more trustworthy. In this work, $n$ usually is larger, for which $P_1(F)/ P_1(\chi^2)$ reduces to several.

Apart from the choice of the statistic, our method for $N=1$ differs from the LS 
one in the average flux being subtracted in the latter (thus yielding 
$n_\|=2$) whereas a constant term is fitted in the former (which can be 
often of significant advantage, see Foster 1995). If the periodic modulation in 
the data differs significantly from a sinusoid (e.g., due to dips, eclipses, 
etc.), then our $N>1$ models account for that more complex shape and perform 
considerably better then the LS one. For example, we show in Section 
\ref{ss:orbhard} that a folded {\it RXTE}/ASM light curve is much better 
described by a model with $N=2$ than $N=1$ with the probability of that 
improvement being by chance of $\sim\! 10^{-13}$.

\subsubsection{Multiple trials}
\label{ss:mdet:mult}

Probability can be assigned to a period found in data according to one of two 
statistical hypotheses. Namely, {\it (i)\/} one knows in advance the trial 
frequency, $\nu_0$ (from other data), and would like to check 
whether it is also present in a given data set or {\it (ii)\/} one searches a 
whole range, $\Delta\nu$, of $N_{\rm eff}$ frequencies and finds the frequency, 
$\nu$, corresponding to the most significant modulation. The two cases 
correspond to the probabilities $P_1$ and $P_{N_{\rm eff}}$ to win in a lottery 
after 1 and $N_{\rm eff}$ trials, respectively, i.e., they represent the false 
alarm probabilities in single and multiple experiments, respectively. They are 
related by $P_{N_{\rm eff}}= 1-(1-P_1)^{N_{\rm eff}}$. Note that 
the hypothesis {\it (ii)\/} and the probability $P_{N_{\rm eff}}$ must be always 
employed in order to claim any new frequency in the object under study. The 
hypothesis {\it (i)\/} is rarely used. However, since $P_1<P_{N_{\rm eff}}$, it 
is the more sensitive one. For this reason, we advocate its use in the 
situations where the modulation frequency is already known, and we aim at 
checking for its manifestation in the same object but in a new band, new data 
set, etc. We stress that we do not use the hypothesis {\it (i)\/} to claim 
any new frequency.

In this work, we use $P_1$ for cases with the orbital periodicity and the $\sim$150-d superorbital period, found before in numerous studies, and use $P_{N_{\rm eff}}$ for all other periodicities. We calculate $P_1$ and $P_{N_{\rm eff}}$ applying all due corrections (see below), and requiring stability of a period after removal of a modulation with another frequency (prewhitening).

An obstacle hampering use of the {\it (ii)\/} hypothesis is that no analytical 
method is known to calculate $N_{\rm eff}$. The number $N_{\rm eff}$ corresponds 
to independent trials, whereas values of periodograms at many frequencies are 
correlated because of the finite width of the peaks, $\delta\nu$, and because of 
aliasing. As no analytical method is known to determine $N_{\rm eff}$, Monte Carlo simulations have been used (e.g., Paltani 2004). Here, we use a simple 
conservative estimate, $N_{\rm eff}= \min(\Delta\nu/\delta\nu, N_{\rm calc},n)$, 
where $N_{\rm calc}$ is the number of the values at which the periodogram is 
calculated. The estimate is conservative in the sense that it corresponds to the 
upper limit on $P_{N_{\rm eff}}$, and thus the minimum significance of 
detection. This effect applies to all methods of period search (Horne \& 
Baliunas 1986). In general, it may reduce significance of a new frequency 
detection for large $N_{\rm eff}$ as $P_{N_{\rm eff}}\gg P_1$. In practice, it 
underscores the role of any prior knowledge, in a way similar to the Bayesian 
statistics: with any prior knowledge of the given frequency we are able to use 
the hypothesis {\it (i)\/} to claim the detection with large significance (small 
$P_1$).

\subsubsection{Correlation length}
\label{ss:mdet:corr}

The $P_1$, and other common probability distributions used to set the detection 
criteria, are derived under the assumption of the noise being statistically 
independent. Often this is not the case, as seen, e.g., in light curves of 
cataclysmic variables (CVs). The correlated noise, often termed red noise, 
obeys different probability distribution than the standard $P_1$,  and hence may 
have a profound effect. For example, noise with a Gaussian autocorrelation 
function (ACF) correlated over a time interval, $\delta t$, yields a power 
spectrum with the Gaussian shape centered at $\nu=0$ and the width 
$\delta\nu=1/\delta t$. It may be demonstrated that the net effect of the 
correlation on $P_1$ in analysis of low frequency processes is to decimate the 
number of independent observations by a factor $n_{\rm corr}$, the average 
number of observations in the correlation interval $\delta t$ 
(Schwarzenberg-Czerny 1991). Effectively, one should use $n_{\perp}/n_{\rm 
corr}$ and $\Theta/n_{\rm corr}$ instead of $n_{\perp}$ and $\Theta$ in calculating $P_1$. This result holds generally, for both least squares and maximum likelihood analyses of time series. In the respect of the red noise, accretion-powered X-ray sources resemble CVs and for them the red noise can also have profound consequences, see e.g., the simulations of Kong et al.\ (2002). In order to take into account this effect here, we check the ACF of the fit residuals.

For independent observations, $m=2$ consecutive residuals have the same sign on 
average (e.g., Fisz 1963). Thus, counting the average length, $m$, of series of 
residuals of the same sign provides an estimate of the number of consecutive 
observations being correlated, $n_{\rm corr}$. Note that $m=n/l$ where $l$ is 
the number of such series (both positive and negative). For correlated observations, the average length of series with the same sign is $m=2n_{\rm corr}$, which allows us to calculate $n_{\rm corr}$.

Let $\Theta$ denote the Fisher-Snedecor statistics from the mhAoV periodogram (i.e. from Fourier series fit) computed for $n_{\|}=2N+1$ parameters, $n$ observations and $n_{\perp}=n-n_{\|}$ degrees of freedom. To account for $n_{\rm corr}$, we calculate $P_1$ as follows,
    \begin{equation}
  P_1=1- F\left(n_{\|},\frac{n_{\perp}}{n_{\rm
corr}};\frac{\Theta}{n_{\rm corr}}\right)=
  I_{z}\left(\frac{n_{\perp}}{2n_{\rm
corr}},\frac{n_{\|}}{2}\right)\label{P1},
\end{equation}
where
\begin{equation}
z= \frac{n_{\perp}}{n_{\perp}+n_{\|}\Theta}\,,
\end{equation}
and $I_z(a,b)$ is the incomplete (regularized) beta function (Abramowitz \& Stegun 1971), see Schwarzenberg-Czerny 1998 and references therein. (In the popular application, {\sc mathematica}, Wolfram 1996, that function is called 
$\mbox{\texttt{BetaRegularized}}$.)

\subsection{The method of fitting folded light curves}
\label{ss:fits}

After establishing the existence of a period, we characterize the
shape of the modulation by fitting the light curve, which we fold
and average using a binning appropriate for the given statistics and the complexity of its shape. The shown errors on the folded and averaged light curves are obtained from the dispersion (rms) of the fluxes pre-averaged over each real-time bin falling into a given phase bin. (The pre-averaging is done because variability over time scales shorter than the time length of a phase bin is unimportant for the analysis.) In fitting the light curves, we take into account the errors obtained in this way. All the uncertainties shown below are $1\sigma$.

We assume the modulation corresponds to a multiplicative factor (rather than to an additive flux component). This is suitable to describe modulation due to phase-dependent absorption, but it also can describe other effects. Thus,
\be
   {\cal F}(t) = {\cal F}_{\rm intr}(t) A_{\rm mod}(t),
\label{ft}
\ee
where ${\cal F}_{\rm intr}(t)$ is the intrinsic flux and $A_{\rm mod}(t)$ is the modulation factor.

Our use of the flux logarithm, $G(t)= \ln {\cal F}(t)$, allows us then to describe the modulation as additive. Then we use nonlinear least-square method to obtain the best-fitting $N$-harmonic Fourier series,
\be
G_{\rm mod}(t) = G_0 - \sum_{k=1}^N G_k \cos[2\upi k \nu (t-T_0) - \varphi_k],
\label{ffunction}
\ee
where $t$ is measured from some initial moment, $T_0$, $G_0$ is a (fitted) constant term, and we use the same value of $N$ as for the corresponding periodogram. The value of $N$ is established by the F-test based on the values of $\chi^2$ of the fit with equation (\ref{ffunction}). Note that we do not treat $\nu$ as a free parameter but instead use either the values obtained from the periodograms or from other data (e.g., the orbital frequency from spectroscopy), compatible with fitting here the folded and averaged light curve. 

The obtained modulation can be removed from the light curve by,
\begin{eqnarray}
\lefteqn{ G_{\rm intr}(t) = G(t) - G_{\rm mod}(t) + G_0\nonumber}\\
\lefteqn{
\qquad\quad\,\,\, =G(t)+\sum_{k=1}^N G_k \cos[2\upi k \nu (t-T_0) - \varphi_k],
\label{pw}}
\end{eqnarray}
which preserves the average flux. On the other hand, if the modulation is caused by absorption, one can correct the light curve for the variable part of the absorption,
\be
   G_{\rm intr}(t) = G(t) - G_{\rm mod}(t) + G_{\rm max},
\label{remove_mod}
\ee
where $G_{\rm max}=\max [G_{\rm mod}(t)]$, which increases the average logarithm of the flux by $(G_{\rm max}-G_0)$, see Tables 2 and 3.

We remove the orbital modulation prior to searching for longer periods (prewhitening), using equation (\ref{pw}), in order to reduce the noise level. We then use the periodogram to identify the frequency of the next most prominent modulation, and fit the Fourier series of equation (\ref{ffunction}). Then we prewhiten it again, i.e., divide the data by the new modulation. We repeat the procedure until the peak amplitude becomes close to the noise level. To test the persistence and significance of the detected modulation, we then repeat the procedure using a half of the data. On the other hand, we remove the main superorbital modulation found by us (see Section \ref{superorbital} below) before calculating final results for the orbital modulation.

We define the relative modulation depth by
\be
\phi_{\rm mod}\ = \frac{{\cal F}_{\rm max}- {\cal F}_{\rm min}} {{\cal F}_{\rm max}},
\label{phimod}
\ee
where ${\cal F}_{\rm min}$ and ${\cal F}_{\rm max}$ are the minimum and maximum flux of the fitted model. This definition, though it differs from the standard peak-to-peak one, is suitable for modulation caused by absorption. For $N=1$, the modulation depth and its standard deviation are,
\begin{equation}
\phi_{\rm mod}= 1-\exp(-2 G_1),\qquad \delta \phi_{\rm mod}=2 \exp(-2 G_1) \delta G_1,
\end{equation}
where $\delta G_1$ is the standard error of $G_1$. For $N>1$, $\phi_{\rm mod}$ is found numerically, and $\delta \phi_{\rm mod}$, by propagation of errors.

\subsection{Spectral windows}

\begin{figure}
\centerline{\includegraphics[height=7.5cm,angle=-90]{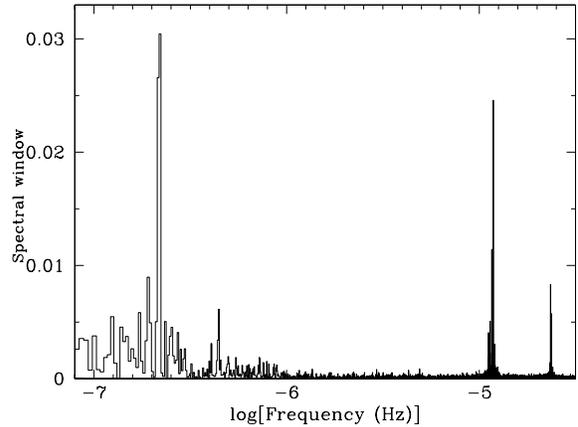}}
\caption{The spectral window of the \asm\ light curve. The peaks at $2.2\times 10^{-7}$ Hz (52.6~d) and $1.2\times 10^{-5}$ Hz (0.96~d) appear to be caused by the precession of \rxte\/ and observation scheduling, respectively.
}
\label{f:asmwin}
\end{figure}

The discrete spectral window (DSW),
\begin{equation}
 |w(\nu)|^2 \! =
 \! n^{-2} \!
 \left[\! \left( \sum_{k=1}^{n} \cos 2\upi\nu t_k \right)^{\!\!2}
 \! \! + \!
 \left( \sum_{k=1}^{n} \sin 2\upi\nu t_k,
 \right)^{\!\!2} \right],
\label{wnu}
\end{equation}
where $t_k$ is the time of the $k$th observation, identifies features related to regularities of the light curve sampling (e.g., Deeming 1975). We have found that only the DSW of the \asm contains interesting features. Namely, there are two strong peaks at 52.6~d and 0.96~d and their harmonics, see Fig.\ \ref{f:asmwin}. We have also calculated the DSW for three other bright X-ray sources, Crab, Cyg~X-3 and GX~5--1, and found the same peaks.

The likely cause of the 52.6-d peak is precession of the \rxte\/ satellite, with period in the range of 50--60 d (A. M. Levine, personal communication). The precession causes a modulation in the frequency of passing through the South Atlantic Anomaly, when the detectors are switched off. The $\sim$1-d feature appears to be due to \rxte\/ scheduling (Zdziarski et al.\ 2004; Farrell, O'Neill \& Sood 2005). A peak of the DSW at a frequency $\nu$ may give rise to aliases of actual modulation at frequency offsets of $\pm \nu$. However, we have found no such features in the studied periodograms.

\section{Orbital modulation}
\label{s:orb}

\subsection{The hard state}
\label{ss:orbhard}

Whenever we find a period of $\simeq 5.6$ d, it is compatible within the measurement errors (Table 2) with the spectroscopic orbital period. Given this lack of evidence of any difference between the period of the modulation and the actual orbital period, we use the best available ephemeris for all the data sets in fitting the modulation shape. We use the period, ${\cal P}$, of Brocksopp et al.\ (1999b) and choose $T_0$ based on LaSala et al.\ (1998), which gives the ephemeris of
\begin{equation}
  \mbox{min[MJD]} = 50234.79(\pm 0.01) + 5.599829(\pm 0.00002)E,
  \label{eph1}
\end{equation}
where $E$ is an integer and $T_0$ was chosen to be as close as possible to all of the intervals of our main data sets.

\begin{figure}
\centerline{\includegraphics[height=8.4cm,angle=-90]{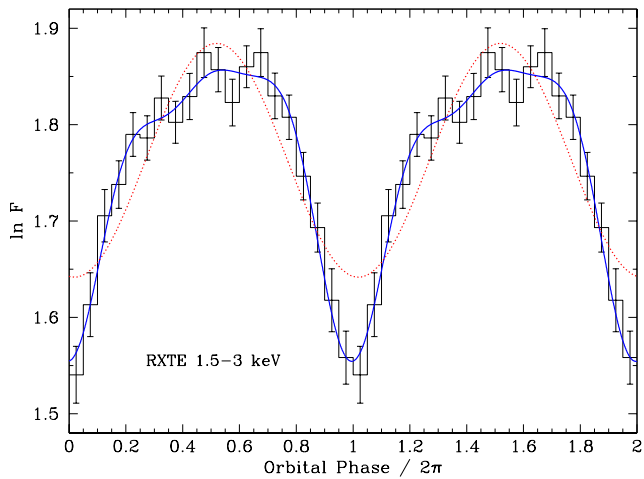}}
\caption{Comparison of fits to the orbital modulation in the 1.5--3 keV range of
the \asm with one (dots) and three (solid curve) Fourier harmonics. We see that the single sinusoid fails to describe the shape of the modulation.}
\label{f:test}
\end{figure}

Following the method of Section \ref{s:ta}, we have found that the orbital modulation has a complex shape for some data sets, for which the sinusoidal description ($N=1$) is clearly insufficient. Using the F-test (e.g., Bevington \& Robinson 1992) for the fits on the rebinned folded light curves, we have found that $N=3$ harmonics of the Fourier series of equation (\ref{ffunction}) are required, in particular, for the {\it RXTE}/ASM data. In the case of the 1.5--3 keV band and for the folded and averaged light curve rebinned into 100 bins, the probability that the chance improvement of the fit from $N=1$ to $N=2$ is $\sim 10^{-13}$, and from $N=2$ to $N=3$, 0.004. Then, $N=4$ is not required, with the probability of 0.67. Fig.\ \ref{f:test} compares the results for $N=1$ and $N=3$. The shape of the orbital modulation can be described as relatively flat for the normalized phase $\sim$0.3--0.7 and with a sharp dip around the phase 0/1.

\begin{figure}
\centerline{\includegraphics[width=8.4cm]{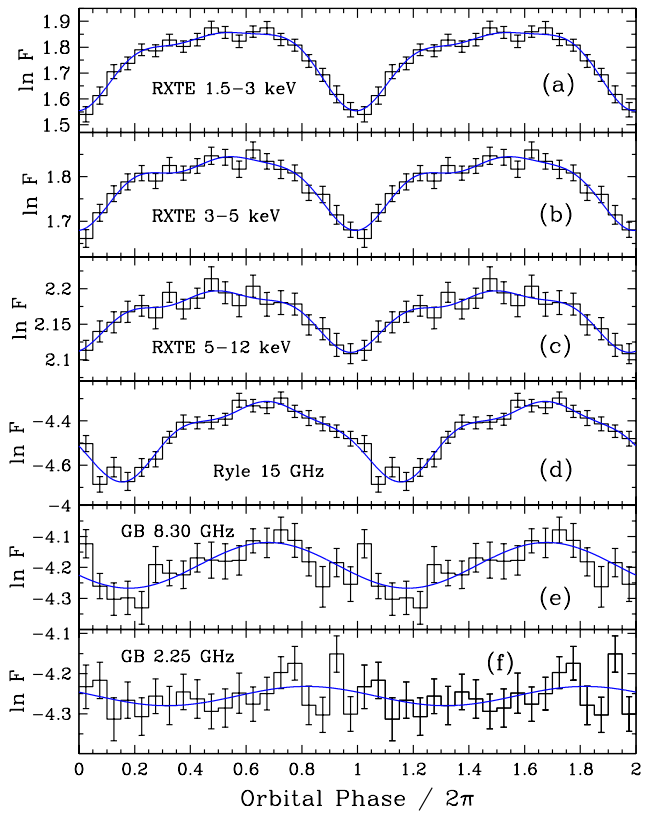}}
\caption{The phase diagrams of the orbital modulation for the \asm and radio data in the hard state. The solid curves represent the fits with $N=3$. See Table 2 for details.
}
\label{f:orb}
\end{figure}

Our results are given in Table 2 (but note that we give the values of $n$ only in Table 3) and Fig.\ \ref{f:orb}. We treat jointly the hard-state sub-intervals of the  {\it RXTE}/ASM and Ryle data (Table \ref{t:lc}). The fractional modulation in the \asm data decreases with the increasing energy, and is in good agreement with the results of W99.

The 15 GHz radio data, see Table 2 and Fig.\ \ref{f:orb}(d), also require $N=3$. Notably, the minimum of the radio modulation is significantly shifted with respect to the zero phase, by 0.153, or 0.86 d. The corresponding ephemeris (using the spectroscopic period) is,
\begin{equation}
  \mbox{min[MJD]} =  50235.65(\pm 0.01) + 5.599829E.
  \label{eph_ryle}
\end{equation}
A similar, but smaller (0.11 or 0.67 d) offset was earlier reported by P99. The difference appears to be due to both the fit with $N=1$ and a shorter light curve (2 yr) used by them. In particular, the fit with $N=1$ for the current data set yields an offset of 0.14.

The fractional modulation depth of the radio emission decreases with the decreasing frequency, see Table 2 and Fig.\ \ref{f:orb}(e--f). The quality of the 8.3 and 2.25 GHz data is also lower than that of the 15 GHz data, and $N=1$ is sufficient to describe the modulation. Interestingly, the offset of the minimum increases with the decreasing frequency, reaching $0.32\pm 0.09$ ($1.8\pm 0.5$ d) at 2.25 GHz. Since $N=1$, the values of the offset at 8.3 and 2.25 GHz follow directly from Table 1. These results are similar to those of P99, who studied a part of those data up to 1998.

For the 20--100-keV BATSE data, we calculate the mhAoV periodogram with $N=1$, which yields $\Theta=3.7$, corresponding to the probability of $P_1\simeq 0.18$, see Table 2. Although this is a very weak significance, the 20--100 keV modulation is probably real given our a priori knowledge of the orbital period and the presence of very significant modulation in softer X-rays. Our results agree with those of Paciesas et al.\ (1997), who found orbital modulation in the 20--300 keV BATSE data. The case for orbital modulation in the 100--300 keV band is still weaker. With the mhAoV method, we still find the period with a small error, ${\cal P}= 5.597\pm 0.005$, in spite of $\Theta=1.0$. Overall, the BATSE data are fully compatible with the presence of (weak) modulation, but they do not prove it. On the other hand, an orbital modulation with parameters similar to those of our best fits is implied by the presence of Compton scattering in the companion wind, see Section \ref{ss:disorb}. 

The \ginga data show the orbital modulation at high significance (Table 2). Our results agree well with those of K00. We note here that the {\it Ginga\/} data clearly show higher depths of the modulation than the {\it RXTE}/ASM ones, especially for the 6--20 keV band. Taken at the face value, the data indicate that the wind from the companion was significantly stronger during 1987--1991 than that after 1996.

Surprisingly, we find no orbital modulation in the periodograms from \vela
and {\it Ariel 5}. This, however, agrees with the results of Holt et al.\ (1979) and PTH83, who reported that the 5.6~d modulation was not significantly detected in the Fourier transforms of those data.

\subsection{The soft state}
\label{ss:orbsoft}

\begin{figure}
\centerline{\includegraphics[height=8.4cm,angle=-90]{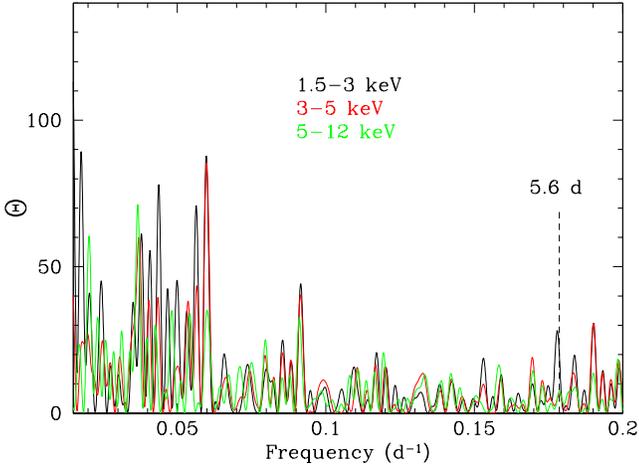}}
\caption{The mhAoV periodograms for the 2001--2002 soft state showing the lack of evidence for orbital modulation in the \asm data.}
\label{f:asmsoft}
\end{figure}

We have analyzed two occurrences of the soft state in the {\it RXTE}/ASM and Ryle data, 2001--2002 and 2003, see Fig.\ \ref{f:lc}. The long duration of the former provides the best available constraints. The mhAoV periodograms for the {\it RXTE}/ASM data are shown in Fig.\ \ref{f:asmsoft}. We see no evidence for orbital modulation, confirming the result of W99 for the 1996 soft state. This is probably due to a strong increase of the degree of wind ionization by the enhanced soft X-ray flux (W99). 

On the other hand, we find orbital modulation in the 15-GHz data in the soft state of 2001--2002, with ${\cal P}=5.6198\pm 0.0043$~d at $P_1=0.012$ and $\phi_{\rm mod}=(24.5\pm 8.5)\%$. This modulation depth is compatible within errors with that in the hard state (Table 2). However, its possible decrease can be due to the decrease of the mass loss rate from the secondary in the soft state by $\sim$20 per cent, G03 (which also contributes to the reduction of the X-ray absoption). The relatively low significance level appears to be due to the weakness and strong aperiodic variability of the soft-state 15-GHz flux (Fig.\ \ref{f:lc}b). Physically, we do expect the modulation to be present as its mechanism is (most likely) free-free absorption (Brocksopp et al.\ 2002), which is insensitive to the ionization level of elements heavier than He (which are in turn responsible for the X-ray absorption).

\section{Superorbital Modulation}
\label{superorbital}

\begin{figure}
\centerline{\includegraphics[width=8.4cm]{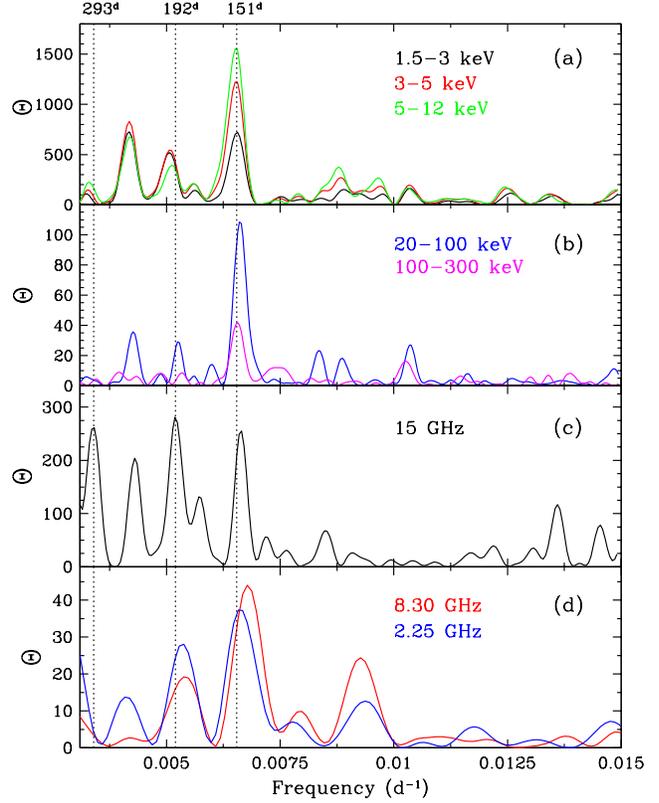}}
\caption{The low-frequency hard-state mhAoV periodograms (after removing the orbital modulation). {\it (a)} The \asm data. {\it (b)} The \batse data. {\it (c)} The Ryle radio data. (d) The GBI radio data. }
\label{f:trf}
\end{figure}

\begin{figure*}
\centerline{\includegraphics[width=12cm]{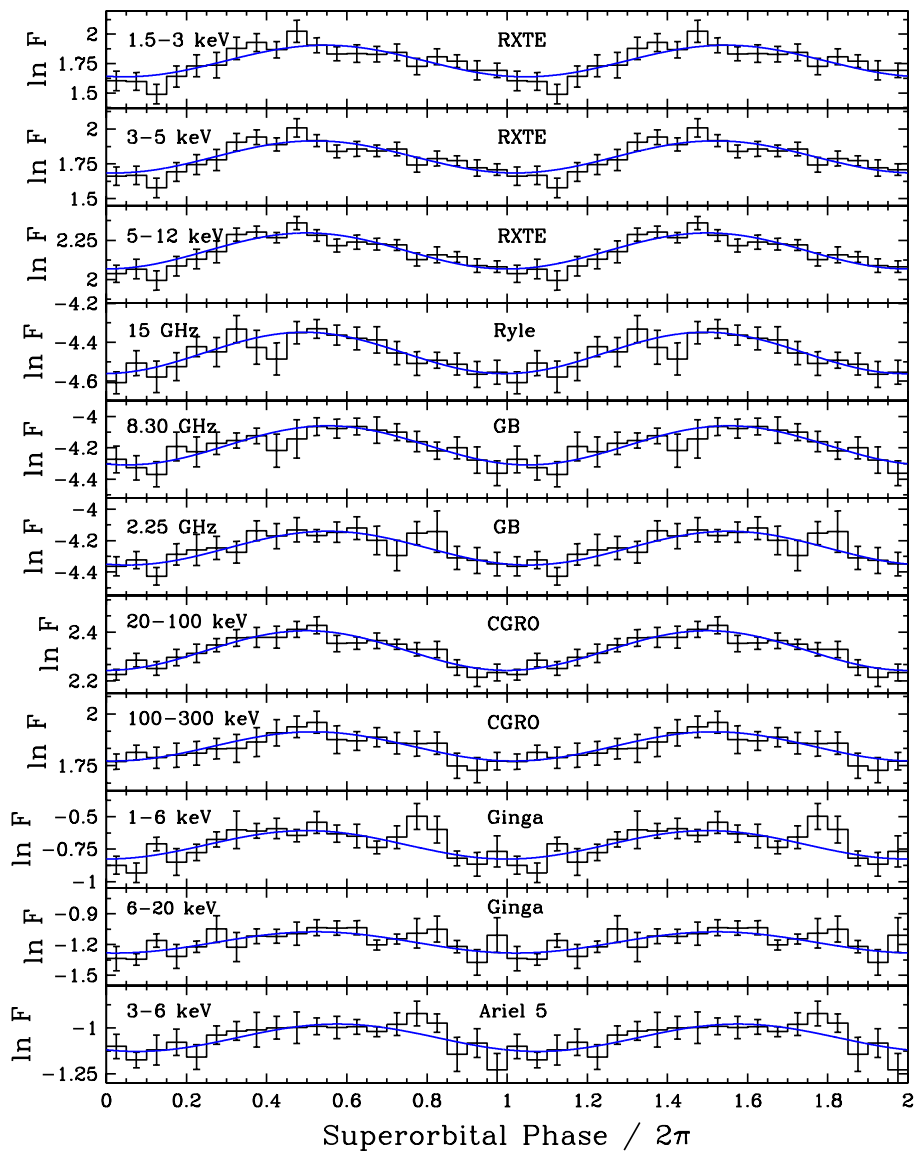}}
\caption{The phase diagrams of the superorbital modulation in the hard state for the ephemeris of equation (\ref{eph}), modelled by sinusoids. Note the approximate constancy of the phase for all the light curves. See Table 3 for the parameters. }
\label{f:pph}
\end{figure*}

Here, we search for variability with periods longer than the orbital one. We use the same method as in Section \ref{s:orb}. However, we find that sufficient fits to the folded and averaged light curves can be obtained with $N=1$. The main results are given in Table 3 and Figs.\ \ref{f:trf} and \ref{f:pph}. 

We have found a $\sim$150~d period to be the most pronounced superorbital one in the hard-state light curves. Note that we remove the orbital modulation, as described in Section \ref{ss:fits} (except for \vela and \ariell, which do not show orbital modulation), and other superorbital modulations (whenever present, see the bottom part of Table 3) prior ot calculating the final results for the $\sim$150~d superorbital modulation. 

We then determine the common ephemeris to all the data using the weighted average, $151.43\pm 0.20$ d, of all the $\sim$150~d periods, and choose $T_0$ (around MJD 50500) based on the {\it RXTE}/ASM 5--12 keV data, which provide the highest signal-to-noise ratio of the modulation (see Table 3). This yields
\be
   {\rm min[MJD]} = 50514.59 + 151.43(\pm 0.20)E.
\label{eph}
\ee
We find that each of the individual periods is consistent with the weighted average within $\la 2\sigma$.

Apart from the ${\cal P}\sim$150~d, we find also some longer periods, $>$200~d, in the earliest observations, i.e., by {\it Vela 5B, Ariel 5, Ginga}. For those, we have arbitrarily chosen $T_0 = 44450.0$ (MJD). We separately discuss the $\sim$290-d period from the \vela 10-yr monitoring in Appendix \ref{a:vela}.

Fig.\ \ref{f:trf}(a) shows the hard-state \asm periodograms. The most significant peak in all three energy bands corresponds to ${\cal P}\simeq 150$~d. Similar values have been earlier reported in the 1.5--12 keV ASM data by B99, Benlloch et al.\ (2001, 2004) and \"{O}zdemir \& Demircan (2001) and in optical data by Karitskaya et al.\ (2001). Also, that periodicity is confirmed by the structure-function analysis, see Appendix B. We find the fractional modulation in the three bands to be compatible with constant, see Table 3.

Prewhitening with the $\sim$150-d modulation has resulted in the lack of other significant periodicities in the {\it RXTE}/ASM data. Although we see two peaks with longer ${\cal P}$ in the mhAoV periodograms, Fig.\ \ref{f:trf}(a), their position change after prewhitening with the orbital and the $\sim$150-d periods. Also, those periods change when we consider shorter intervals of the ASM data. In particular, we do not confirm here the 182.5-d period claimed by \"{O}zdemir \& Demircan (2001).

Fig.\ \ref{f:trf}(b) shows the  20--100 keV and 100--300 keV mhAoV periodograms corresponding to the hard state observed by BATSE. The only significant peak corresponds to ${\cal P}\sim150$ d. It is also confirmed by the structure-function analysis, see Appendix B. Our findings contrast those of Paciesas et al.\ (1997), who reported no periodicities at frequencies $\la 0.1$ \dayys. Still, one can see a Fourier peak corresponding to $\sim 150$~d in their power spectrum. Then, B99 found ${\cal P}\simeq 142.0\pm 7.1$~d in the 20--100 keV data before the 1996 state transition, in agreement with our findings. Interestingly, the depth of the modulation is lower than that of the 1.5--12 keV emission, see Table 3.

The 15-GHz hard-state data show ${\cal P} =150.7\pm 0.6$~d, see Fig.\ \ref{f:trf}(c). However, that peak is accompanied by another statistically significant periodicity at $192.3\pm 1.0$~d. We have checked that the $\sim 150$~d periodicity remains after prewhitening with the ${\cal P}\simeq 192$~d period. The 15-GHz data also show ${\cal P}\simeq 293$~d period, similar to the $\sim$294~d X-ray modulation found by PTH83 in the {\it Vela 5B\/} data. However, prewhitening of the radio light curve with both the $\sim$150 d and and 192~d periods causes the change of that longest period from 293~d to $\sim$350~d, which casts doubt on the reality of that modulation. In fact, neither the 192 nor 293-d periodicity appear applying the structure-function analysis, see Appendix B.

The GBI 8.3 and 2.25 GHz data also show prominent modulation at $\sim$150 d, see Table 3 and Figs.\ \ref{f:trf}(d) and \ref{f:pph}. Their depth is compatible with being equal to that of the 15 GHz emission. Interestingly, the periodograms also show peaks around $\sim$190 d, similar to those seen in the Ryle data. Subsets of the Ryle and GBI data up to 1998 were earlier analyzed by B99, who obtained results compatible with ours.

The \gingaa\ data were earlier analyzed by K00. They reported two periods, 150 d and $\sim 210$--230 d. Here, we confirm their results, obtaining two peaks at
151 d and 230 d.

In the {\it Ariel 5\/} 3--6 keV hard-state data, we have found two prominent peaks (not shown here). The first one at ${\cal P}\simeq  278.6\pm 3.1$~d is also confirmed by the structure-function analysis (Appendix B). It is close to the ${\cal P}\simeq 296\pm 10$~d reported by PTH83. 

The second peak corresponds to ${\cal P}\simeq 151.3\pm 0.8$~d, which remains significant after prewhitening with the first period. We note that the $\sim 150$~d periodicity is seen, in fact, in the Fourier spectrum of PTH83.

Remarkably, all the $\sim$150-d modulations since 1976 are consistent with not only having the same period, but also show an almost constant phase, see Table 3 and Fig.\ \ref{f:pph}. The phase constancy also support our determination of the period from the weighted average, as an error on that would result in phase shifts in data separated by long intervals.

The approximately constant phase is kept in spite of a number of soft-state occurrences interrupting the hard state. The longest available soft-state period, of 2001--2002 is still too short (391 d, Table \ref{t:lc}) to enable us to study the superorbital modulation in that state itself.

\section{Discussion}
\label{s:discussion}

\subsection{Orbital modulation}
\label{ss:disorb}

We consider the usual spherically symmetric approximation to the radial density profile of the wind from the companion,
\be
   n(R) = \left( \frac{R_\star}{R} \right)^2 \frac{n_0}{[1-(R_\star/R)]^\alpha}
\label{nrprofile}
\ee
(e.g., Castor, Abbott \& Klein 1975), which corresponds to the velocity profile of
\be
   v(R) = v_\infty \left( 1 - \frac{R_\star}{R} \right)^\alpha,
\ee
and where $R_\star$ is the radius of the star, $n_0 = \dot{M}_{\rm loss}/(m_{\rm H} 4\upi R_\star^2 v_\infty)$ is the density parameter, $\dot{M}_{\rm loss}$ is the mass-loss rate, $m_{\rm H}$ is the hydrogen mass, $v_\infty$ is the terminal velocity of the wind, and $\alpha$ is the power-law index of the dependence.

These formulae were used by W99, who have successfully explained the orbital modulation seen in the hard state in their {\it RXTE}/ASM data by phase-dependent absorption in the ionized wind. They found $i\simeq 30^{+10}_{-20}\degr$ as the inclination implied by the modulation profile.

\begin{figure}
\centerline{\includegraphics[height=8cm,angle=-90]{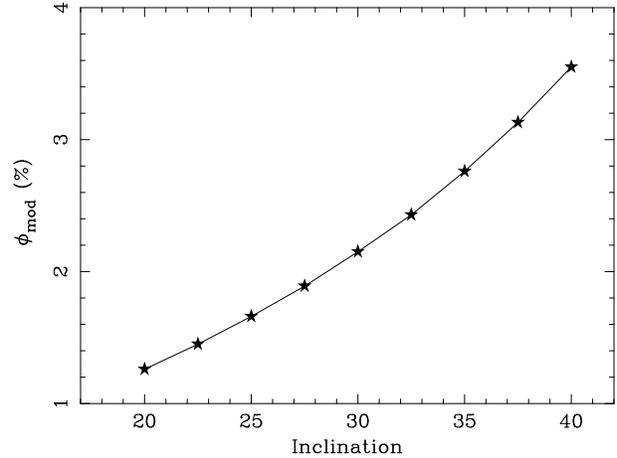}}
\caption{The fractional orbital modulation due to Thomson scattering in the wind as the function of the inclination, see Section \ref{ss:disorb}.
}
\label{f:mod}
\end{figure}

Here, we consider modulation due to Thomson scattering by the wind, relevant to the BATSE 20--100 keV energy range. We obtain then the fractional modulation depth [eq.\ (\ref{phimod})] of
\be
   \phi_{\rm mod} = \frac{\exp[-\tau_{\rm T}(\upi)]
   - \exp[-\tau_{\rm T}(0)] }
   { \exp[-\tau_{\rm T}(\upi)]  },
\label{phimod2}
\ee
where $\tau_{\rm T}$ is the Thomson optical depth integrated along the line of sight from the black hole through the wind,
\be
  \tau_{\rm T}(\phi_{\rm orb}) = \sigma_{\rm T}
  \int_0^\infty n[R(r,\phi_{\rm orb},i)] {\rm d}r
\label{tauth}
\ee
and $\sigma_{\rm T}$ is Thomson cross section. We have assumed the rate of mass
loss by the companion during the hard state of $\dot{M}_{\rm loss}\simeq 1.6\times 10^{20}\,{\rm g\,s}^{-1}$ (G03), $R_\star \simeq 1.58\times 10^{12}\,{\rm cm}$, the binary separation, $a = ({\cal P}/2\upi)^{2/3} [G(M_1+M_2)]^{1/3}$, of $2.27 R_\star$ (Zi\'o{\l}kowski 2005), and $v_\infty \simeq 1586\,{\rm km\,s}^{-1}$, $\alpha=1.05$ (Gies \& Bolton 1986). Then, $n_0 \simeq 1.9\times 10^{10}$ cm$^{-3}$. The results are shown in Fig.\ \ref{f:mod}. The observed orbital modulation in the Thomson regime (20--100 keV) of $\sim$3 per cent corresponds to $i\sim 35\degr$, and is in agreement with the corresponding result of W99. These theoretical results, based on optical data, support the reality of the orbital modulation in the BATSE data. We caution, however, that the actual wind in Cyg X-1 is clearly not spherically symmetric, and thus those determinations of $i$ bear an additional systematic error.

The radio modulation is due to free-free absorption of the jet emission in the wind (Brocksopp et al.\ 2002\footnote{Note, however, apparent errors in their equations (2), (6), (8).}). The phase shift of the modulation with respect to the orbital phase is likely due to the time it takes for the jet material, ejected close to the black hole, to propagate to the place where the radio emission originates (i.e., jet bending).

\subsection{Superorbital modulation}
\label{ss:dissuper}

A generally accepted interpretation for the superorbital modulation, observed also in a number of other binary X-ray sources (e.g., SS 433, Her X-1, LMC X-3, LMC X-4), is accretion disc precession (see e.g., Wijers \& Pringle 1999 for a review and references).

\begin{figure}
\centerline{\includegraphics[width=8cm]{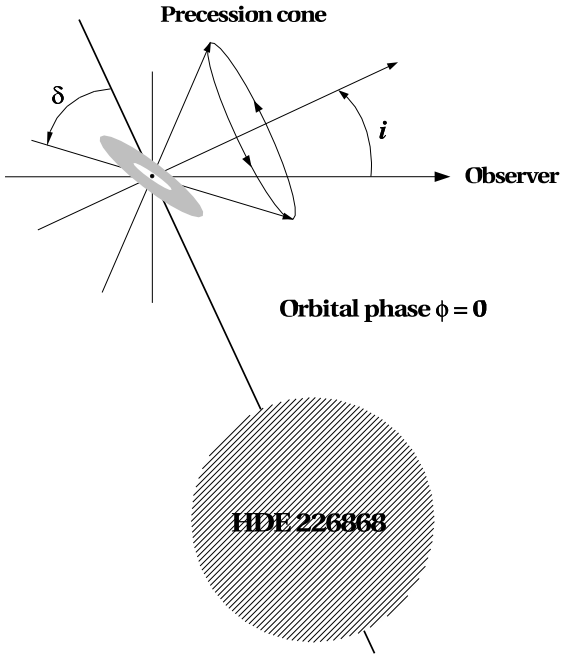}}
\caption{Schematic view of Cyg X-1 at the zero orbital phase. As before, $i$ is the orbital inclination, and $\delta$ is the angle between the inclined disc and the orbital plane.}
\label{geometry}
\end{figure}

If, due to some reason, the disc is inclined with respect to the orbital plane (see Fig.\ \ref{geometry}), it will retrogradely precess due to the tidal forces exerted by the secondary (Katz 1973). This appears to explain the superorbital periodicities in Her X-1, SS 433 and LMC X-4 (Gerend \& Boynton 1976; Leibowitz 1984; Heemskerk \& van Paradijs 1989). The period of the tidally-forced precession is given by,
\begin{equation}
 \frac{{\cal P}_{\rm orb}}{{\cal P}_{\rm sup}}   \simeq \frac{3}{7} \left(\beta R_{\rm L}\over a \right)^{3/2} \frac{\mu}{(1+\mu)^{1/2}} \cos\delta,
\label{cosdelta}
\end{equation}
where $\mu$ is the ratio of the mass of the secondary to that of the compact object, $R_{\rm L}$ is the Roche lobe radius of the primary, $\beta$ is the ratio of the outer disc radius to $R_{\rm L}$, and $\delta$ is the inclination of the disc with respect to the orbital plane (Larwood 1998). Equation (\ref{cosdelta}) assumes that the disc precesses as a rigid body, which is the case when the sound crossing time in the disc is much less than the precession time scale (Papaloizou \& Terquem 1995). We note that the numerical coefficient above depends on the assumed disc model, and may be different from 3/7 (e.g., Larwood 1997).

The Roche lobe radius is well approximated by,
\begin{equation}
{R_{\rm L}\over a} \simeq  \frac{0.49 }{0.6  + \mu^{2/3}\ln (1+\mu^{-1/3})}
\label{r_a}
\end{equation}
(Eggleton 1983). At $2\leq\mu \leq 3$ estimated for Cyg X-1 (Gies \& Bolton 1986; G03; Zi\'o{\l}kowski 2005), $R_{\rm L}/a\simeq 0.32$--0.29. Then,
\begin{equation}
R_{\rm L}\simeq 10^{12} \left( M_1+M_2\over 50\,\msun\right)^{1/3}\, {\rm cm}.
\label{R_L}
\end{equation}

In the case of accretion via Roche-lobe overflow and for thin and weakly viscous discs (Paczy\'nski 1977), $\beta\simeq 0.68$--0.66 at $2\leq\mu \leq 3$. Then for viscous Roche-lobe flows, the size of the disc corresponds to the tidal radius, for which $\beta\simeq 0.87$ (Papaloizou \& Pringle 1977). However, Cyg X-1 accretes via focused wind, in which case the disc size is usually smaller (e.g., Frank, King \& Raine 2002). Therefore, equation (\ref{cosdelta}) puts rather weak constraints on the amplitude of precession, $\delta$. At $\mu=2$, $\delta\simeq 0\degr$--$60\degr$ for $\beta\simeq 0.55$--0.87, and at $\mu=3$, $\delta\simeq 0\degr$--$63\degr$ for $\beta\simeq 0.52$--0.87. Still, equation (\ref{cosdelta}) shows that the precession in Cyg X-1 is fully consistent with being driven by the tidal force of the companion.

On the other hand, Wijers \& Pringle (1999) have considered precession due to radiation-induced warping in X-ray binaries. This can yield either retrograde or prograde precession. The corresponding precession period is roughly given by (see their eq.\ 18),
\begin{eqnarray}
\lefteqn{ {\cal P}_{\rm warp}\sim 2000\left(\alpha\over 0.1\right)^{-{4/5}}\! \left( \epsilon\over 0.1\right)^{-1}\nonumber}\\
\lefteqn{\quad\times
\left(M_1\over 10 \,\msun\right)^{3/4} \left( \dot M\over 10^{17}\,{\rm g\, s}^{-1}\right)^{-{3/10}}\! \left(\beta R_{\rm L}\over 10^{12}\,{\rm cm} \right)^{3/4} {\rm d},}
\label{warping}
\end{eqnarray}
where $\dot M$ is the accretion rate, $\alpha$ is the viscosity parameter and $\epsilon$ is the accretion efficiency. For Cyg X-1,
\begin{eqnarray}
\lefteqn{
\dot M\simeq 2\times 10^{17} { \langle F\rangle \over 4\times 10^{-8}\,{\rm erg\,cm}^{-2}\,{\rm s}^{-1}}  \nonumber}\\
\lefteqn{\qquad\times
\left( d\over 2\,{\rm kpc}\right)^{-2} \left( \epsilon\over 0.1\right)^{-1} {\rm g\, s}^{-1},}
\label{Mdot}
\end{eqnarray}
where $\langle F \rangle$ is the average bolometric hard-state flux and $d$ is the distance, and we used above their values of Zdziarski et al.\ (2002) and Zi{\'o}{\l}kowski (2005), respectively.

The above estimates yield ${\cal P}_{\rm warp}\simeq 10^3$ d, much longer than that observed. We note that Wijers \& Pringle (1999) have given an estimate of ${\cal P}_{\rm warp}\simeq 180$ d for Cyg X-1, but assuming $\alpha$ and $\dot M$ as high as 1 and $1.4\times 10^{18} {\rm g\, s}^{-1}$, respectively. Thus, we consider radiation warping as an unlikely cause of the precession of Cyg X-1. Furthermore, the observed precession has a remarkably stable period, changing at most weakly over $\ga 15$ years, whereas Wijers \& Pringle (1999) note that their mechanism is unlikely to give a stable period for sources with variable luminosity.

Regardless of the mechanism causing the precession, there is the issue of the mechanism causing the flux modulation during the precession. In principle, there are a number of possibilities. One is that the outer edge of the disc (completely optically thick) partially covers the X-ray source. This, however, would require extreme fine-tuning to achieve a $\sim$25 per cent depth of the modulation. Namely, the X-ray source has the size $\sim 10^2 R_{\rm g}$ (where $R_{\rm g}\equiv GM/c^2$), as indicated by the X-ray power spectrum extending to high frequencies and agreement with theoretical prediction on the range of radii where most of the accretion power is released. On the other hand, the outer edge of the disc is at a much larger distance, $\ga 10^5 R_{\rm g}$ (see the discussion above). Another possibility is that the outer part of the disc fully obscures the X-ray source, but we see X-rays scattered in a large-size corona above the disc. This, however, would dramatically affect the X-ray power spectrum, resulting in a cutoff above $\sim$0.1 Hz, which is clearly not seen.  Then, bound-free absorption in a spatially-extended, moderately optically-thin, medium associated with the outer regions of the disc appears to be ruled out as there is rather weak energy dependence of the modulation. On the other hand, a viable scenario is the outer disc wind/corona being almost fully ionized, with scattering away from the line of sight being responsible for the superorbital modulation.

Yet another possibility is that the X-ray emission is intrinsically anisotropic. Such a possibility was considered by B99, who relied in calculating the implied inclination on the blackbody-type anisotropy of a slab with the flux proportional to the projected area. However, there is overwhelming evidence that the dominant radiative process producing X-rays in the hard state of Cyg X-1 (and other black-hole binaries) is thermal Comptonization (e.g., Zdziarski \& Gierli\'nski 2004). The anisotropy of thermal Comptonization (Poutanen \& Svensson 1996) in planar geometries, though substantially weaker than that of the constant specific intensity case, appears fully capable to explain the observed superorbital modulation (work in preparation).

We note that our result on the stability of the superorbital period over $\ga$15 yr rules out the model for strong outbursts of Cyg X-1 by Romero, Kaufman Bernando \& Mirabel (2002). Those outbursts were observed (Stern, Beloborodov \& Poutanen 2001; Golenetskii et al.\ 2002, 2003) on MJD 49727, 49801, 51287, 51289, 52329, 52364 and 52682. Romero et al.\ (2002) have proposed that the events are caused by precession of the jet of Cyg X-1, with the flares corresponding to the strongly beamed emission of the jet crossing our line of sight every precession period. They considered, in particular, the flares on MJD 49727, 49801, 52329, reported by Golenetskii et al.\ (2002). The first two are separated by 74 days, which is about a half of the precession period of Cyg X-1. Romero et al.\ (2002) proposed that that precession period undergoes occasional changes, which would explain the value of 74 d. However, our present results show a remarkable stability of the precession period over the last several tens of years and rule out that explanation. Furthermore, the larger number of the dates of the outbursts compiled by Golenetskii et al.\ (2003) do not show any regularity. Interestingly, the event on MJD 52329 took place at the minimum of $\sim$150~d superorbital cycle.

With the {\it RXTE}/ASM data, we have also looked for signals corresponding to the beat between the orbital and superorbital frequencies. If the physical cause of the superorbital frequency is modulation of the X-ray flux emitted in a given direction by disc precession, the photons reflected (e.g., Magdziarz \& Zdziarski 1995) from the surface of the secondary will change at the beat frequency of the two modulations. Then, the position of a possible peak in the periodogram at either $\nu_{\rm orb}+\nu_{\rm sup}$ or $\nu_{\rm orb}-\nu_{\rm sup}$ would tell us whether the precession is retrograde or prograde. We have thus searched for the corresponding peaks at the periods of 5.40 d and 5.82 d.

\begin{figure}
\centerline{\includegraphics[width=8.4cm]{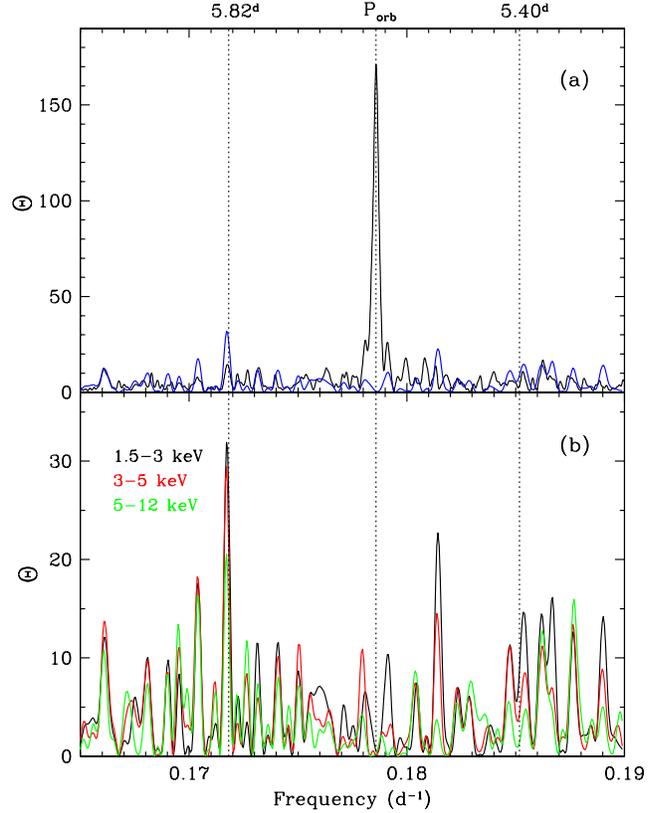}}
\caption{(a) The mhAoV periodograms for the {\it RXTE}/ASM 1.5--3 keV data after prewhitening with both the orbital and superorbital periods (blue curve) and before it (black curve). Both periodograms show peaks corresponding to the period of 5.82 d, possibly  indicating the prograde character of the precession. (b) Detailed shape of the periodograms after the prewhitening in the three ASM bands.
}
\label{f:beat}
\end{figure}

Fig.\ \ref{f:beat}(a) shows the periodograms for the 1.5--3 keV band before and after prewhitening with both the orbital (calculated for $N=3$) and superorbital frequencies. After the prewhitening, the use of $N=1$ is sufficient to account for the shape of any remaining features. We clearly see a pronounced peak at 5.82 d. Fig.\ \ref{f:beat}(b) shows that such peaks are present in all three ASM bands. The statistical significance of the features, $P_{N_{\rm eff}}$, and the relative modulation depth, $\phi_{\rm mod}$, for the 1.5--3, 3--5 and 5--12 keV bands is $1\times 10^{-9}$ and 0.05, $2.2\times 10^{-8}$ and 0.03, $2\times 10^{-5}$ and 0.01, respectively. No statistically significant 5.40-d variability is detected. On the other hand, there is a rather strong peak (though much weaker than that at 5.82 d) at $\sim$5.5 d in the 1.5--3 keV band, which origin is unclear.

This may indicate that the precession in Cyg X-1 is prograde. However, the interpretation of the beat frequency as due to variable reflection from the companion surface presents some problems. Taking the determination of the binary parameters of Zi{\'o\l}kowski (2005), we find that the companion subtends a solid angle of $\sim 0.05\times 4\upi$ as seen from the black hole. The observed modulation amplitude of $\sim$0.05 in the 1.5--3 keV band would require a unit albedo and such precession amplitude that the X-ray source seen from the star surface is fully obscured by the disc. Furthermore, the albedo has to decrease by a factor of several in the 5--12 keV band. We have also checked the relative phase between the orbital and beat modulations. In the above interpretation, the two modulations should be in phase around the minimum of the superorbital modulation, when the disc is most inclined towards the companion. However, we find the ephemeris of the beat modulation to be,
\begin{equation}
  \mbox{min[MJD]} = 50239.70 + 5.82E,
  \label{eph_beat}
\end{equation}
which corresponds to the orbital and beat modulations being in phase instead around the midpoint between the minimum and the maximum of the superorbital modulation. Such relative phases appear to have no geometrical interpretation.

We have also searched for signatures of nutation. This effect is due to a perturbation of the disc rotation at the maximum of the torque exerted on the disc by the companion. This takes place when the star is closest to the point on the outer edge of the disc at the maximum distance from the orbital plane. Therefore, the nutation frequency is twice the beat frequency. We have not found signatures of nutation in any of our data.

We would like to stress the remarkable stability of the $\sim$150-d modulation. This modulation is clearly present starting with the {\it Ariel 5\/} data used by us, i.e., the beginning of 1976, and remains until at least mid 2003. This corresponds to about 65 precession periods. Furthermore, the phase of the modulation has also remained approximately constant, see Table 3. This appears to present a problem for the interpretation of the prograde character of precession. Precession can be prograde due to radiation-induced warping (Wijers \& Pringle 1999), but it is then unlikely to have such constant period given the variable intrinsic luminosity of Cyg X-1 and its state transitions. Superhump-type precession is prograde, but it can take place only for $\mu \ll 1$ (e.g., Frank et al.\ 2002). Then, our results may indicate the presence of a new mechanism causing stable prograde precession at $\mu>1$.

On the other hand, the $\sim$150-d period is not present in the {\it Vela 5b\/} data, $\sim$1969--1979, where, instead, the $\sim$290 d period appears (Appendix \ref{a:vela}, PTH83). Interestingly, we may see a slow evolution of this period in a following time interval, with secondary superorbital periods in the {\it Ariel 5\/} and {\it Ginga\/} data of 278~d and 230~d, respectively. (The change of the long period from $\sim$290 d to a half of this value was noticed by B99.)

\section{Conclusions}
\label{s:conclusions}

We have studied periodic long-term variability of Cyg X-1 using a statistical 
method different from those used before, and apply it, for the first time, to 
the data set spanning over $\sim$30 years. In our method, we calculate the 
periodograms and probabilities with the multi-harmonic analysis of variance, and 
then fit the folded and averaged light curves. We have analyzed X-ray monitoring 
data from {\it Vela 5B, Ariel 5, Ginga, CGRO\/} and {\it RXTE}, and radio data 
from the Ryle and Green Bank telescopes. We have confirmed and refined previous 
results on the 5.6-d orbital modulation of X-ray and radio emission in the hard 
state, caused by the attenuation in the wind of the companion supergiant being 
dependent the binary orbital phase. In particular, we show the detailed 
non-sinusoidal shape of the orbital modulation of X-rays observed by the {\it 
RXTE}/ASM and of the 15 GHz emission. For the first time, we find an orbital 
modulation at 15 GHz in the soft state. We confirm the presence of orbital 
modulation in the 20--100 keV {\it CGRO}/BATSE data.

Then, we find the presence of a $\sim$150-d superorbital period in all of the 
data since $\sim$1976; in particular, we find its presence for the first time in the {\it Ariel 5\/} data. Very remarkably, we find both that period and its phase have been stable over $>65$ superorbital cycles. This coherence poses severe restrictions on any underlying clock mechanism. We find the cause of the 
superorbital modulation to be compatible with accretion disc precession, probably due to the tidal forces exerted by the secondary. Furthermore, we find modulation at the frequency corresponding to the beat between the orbital modulation and the prograde disc precession. On the other hand, we confirm the presence of a $\sim$290-d periodicity in the 1969--1979 {\it Vela 5B\/} data. This indicates a switch from that precession period to its first harmonic around $\sim$1980.

\section*{ACKNOWLEDGMENTS}

This research has been supported by the KBN grants 1P03D01827, 1P03D02529, 1P03D01128, 4T12E04727. The Ryle Telescope is supported by PPARC. We thank Z. Ko{\l}aczkowski and M. Niko{\l}ajuk for help in software. We also thank K. Jahoda, A. Levine, J. Lochner, A. Olech, M. Revnivtsev, A. Szostek, L. Wen, R. Wijers and P. \.Zycki for valuable discussions. The Green Bank Interferometer is a facility of the National Science Foundation operated by the NRAO in support of NASA High Energy Astrophysics programs.

\appendix

\section{The $\mathbf{\sim}$290-day periodicity in the \textbfit{Vela 5B}/ASM data}
\label{a:vela}

The earliest reported long-term periodicity found in Cyg~X-1 data is the $294 \pm 4$~d one found by \vela in the 3--12 keV range (PTH83). The same period was
shortly after found in optical photometric data by Kemp et al.\ (1983). In Section \ref{superorbital}, we have found a similar period of $279\pm 3$ d in the 3--6-keV {\it Ariel 5}/ASM data, as well as a tentative $\sim 293$-d period in the 15 GHz radio data. However, that period has not been confirmed in any other X-ray data.

\begin{figure}
\centerline{\includegraphics[height=7.5cm,angle=-90]{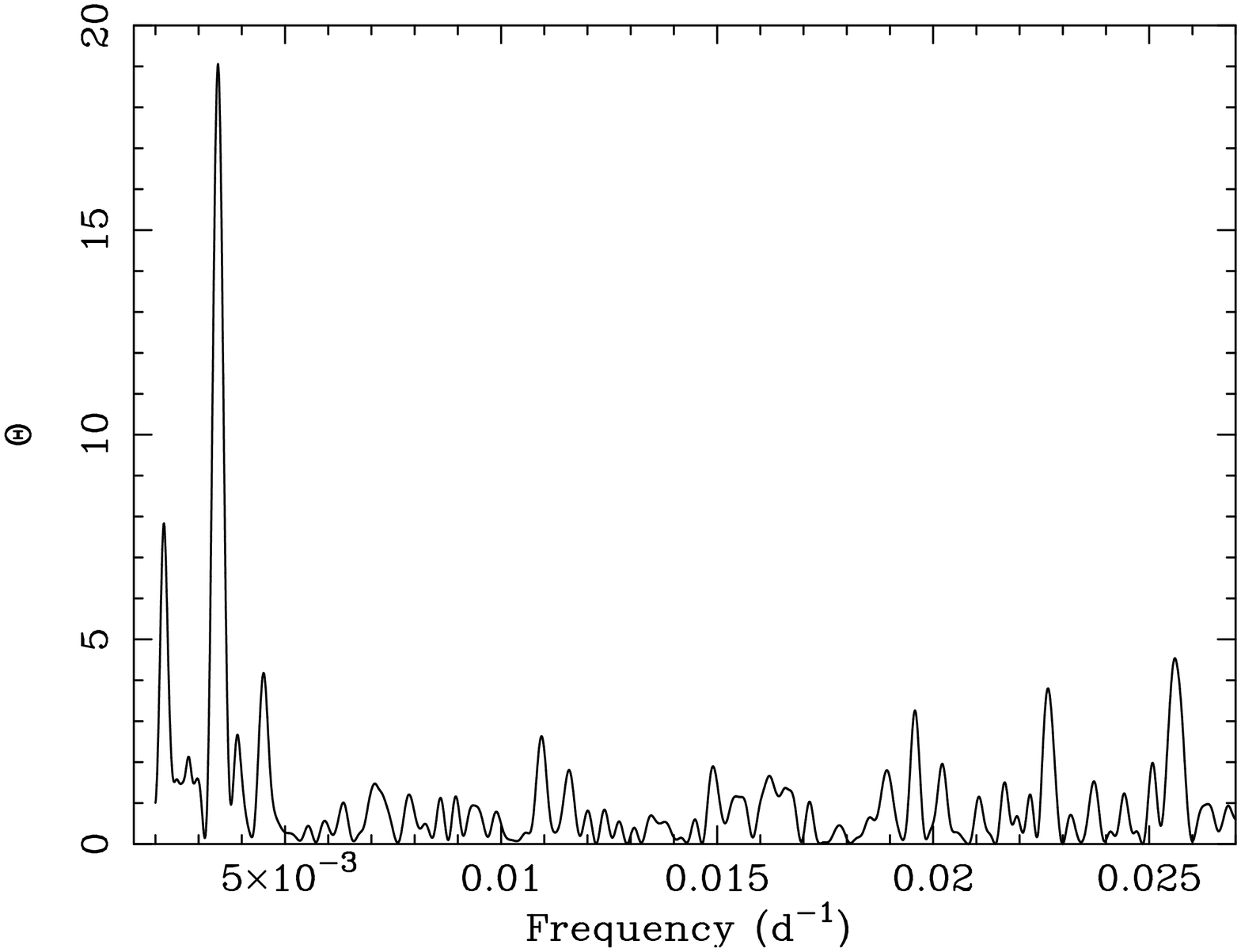}}
\caption{
The mhAoV periodogram for the \vela 3--12 keV light curve. The highest peak is at  $289.5 \pm 1.7$~d period.}
\label{f:vela}
\end{figure}

Here, we analyze the \vela observations of Cyg X-1 binned to 0.25 day averages using  the mhAoV method. The resulting periodogram is shown in Fig.\ \ref{f:vela}. We find a very strong dominant peak at the period of $289.5 \pm 1.7$ d and the absence of any significant feature at the first harmonic. The significance, phase, and the depth of the modulation are given in Table 3.
We have also rebinned the data to 30-d bins, as done by PTH83, and found the maximum slightly shifts to $294\pm 3$ d. Thus, our results are in complete agreement with those of PTH83.

We should note that the precession period of {\it Vela 5B\/} is $\sim$300 d, see the \vela Calibration Guide of HEASARC \footnote{heasarc.gsfc.nasa.gov/docs/heasarc/caldb/docs/vela5b}. This could, in principle, suggest an instrumental origin of the observed periodicity. In order to test this possibility, we have obtained \vela periodograms for three other bright X-ray sources, 4U~0614+09, Cyg X-3 and Her X-1, using the same time span (see Table 1), energy band and binning as for Cyg X-1. In all of those sources, we have found no significant periodicities around $\sim$300 d. These results as well as the confirmation of the $\sim$290-d period in the optical data (Kemp et al.\ 1983) appear to confirm the reality of the peak in the \vela periodogram.

To further test the origin of the observed periodicity, we have generated a Monte-Carlo light curve assuming the count rate equal to the average one of Cyg X-1 with a Gaussian distribution of the errors at the signal-to-noise ratio matching the data at the times matching those of the observations. We have then
calculated the mhAoV periodogram and found no significant $\sim$300-d or other flux periodicity. In fact, $\Theta<5$ was found in the 0.003--0.2 d$^{-1}$ frequency range. This further confirms the reality of the presence of the $\sim$290-d period in the data.

We have also examined the periodograms for the time interval when Cyg X-1 was observed simultaneously by {\it Ariel 5}/ASM and {\it Vela 5B}/ASM, i.e., MJD 42830--44042. We have found those periodograms have the strongest peaks at the periods of 302 d and 291 d, respectively. There is a $\sim$150-d peak in the {\it Ariel 5\/} data, but only as the third strongest one, and no corresponding peak is present in the {\it Vela 5B\/} data. This further supports our findings above.

It then appears that the superorbital period of Cyg X-1 had indeed changed over a time scale $>10$ yr from $\sim$290-d to approximately its first harmonic, $\sim$150 d, seen in all the observations after those by {\it Vela 5B}. During a  transitionary time interval, both periods were present, see Table 3.

\section{The structure function of Cyg X-1}

A method to quantify long-term variability alternative to periodograms is to calculate the structure function (SF). It was introduced by Kolomogorov (1941a, b), and applied for the first time in astronomy by Simonetti, Cordes \& Heeschen (1985), and then used by many authors, e.g., Hughes, Aller \& Aller (1992), Collier \& Peterson (2001), and Czerny et al.\ (2003).

 It is a function of the time domain, with the first-order SF defined by
\be
   {\rm SF}(\tau) = \langle [F(t+\tau)-F(t)]^2 \rangle,
\label{sf}
\ee
where $F(t)$ is the light curve. If $F(t)=A+\sin t$, i.e., it is sinusoidal with the $2\upi$ period and we measure it between the times $a$ and $b$, the structure function can be calculated to be,
\begin{eqnarray}
\lefteqn{ {\rm SF}(\tau)
 =  {1\over b-\tau -a}
            \int_a^{b-\tau} [\sin(t+\tau)-\sin t]^2 \ {\rm d}t \label{sfsin}} \\
\lefteqn{\qquad\quad =
  2\sin^2\frac{\tau}{2} \left[1-\frac{2\cos(a+b)\sin(a-b+\tau)}{b-a-\tau} \right],}
           \label{sfint}
\end{eqnarray}
where $\tau<b-a$, and the second (boundary) term in brackets is negligible for $\tau\ll b-a$. We see that the structure function has {\it minima\/} (at 0) for $\tau$ equal to the period and its subharmonics, i.e., $\tau=2k\upi$, where $k$ is an integer. This should be bear in mind when analyzing results of application of this technique to observed light curves.

\begin{figure}
\centerline{\includegraphics[width=8.5cm]{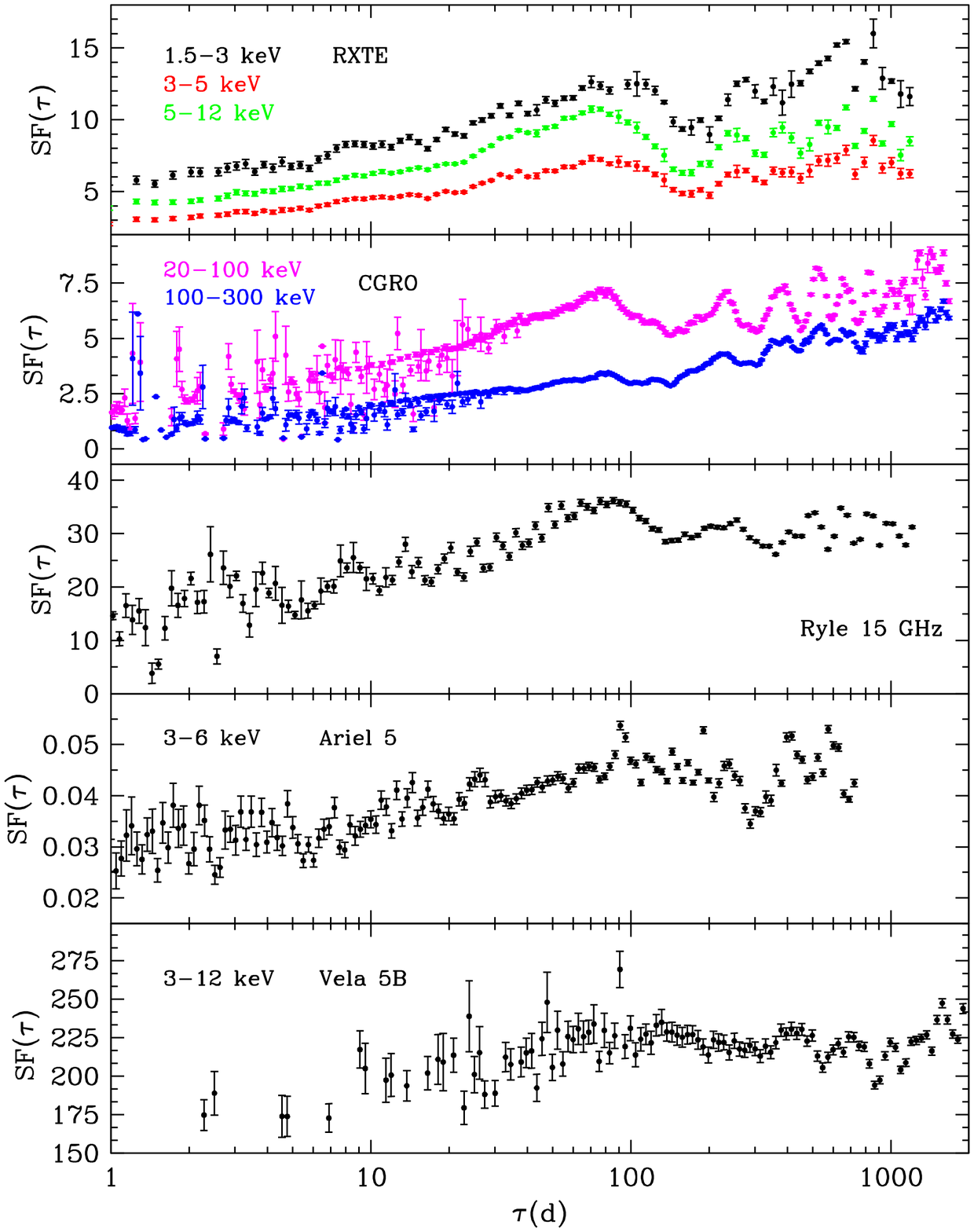}}
\caption{The structure function for our hard-state light curves. Periodic variability in the data manifests itself as dips at $\tau$ equal to the period and its subharmonics.
}
\label{f:sf}
\end{figure}

We then search for long-term periodicities in our light curves corresponding to the hard state (except that of {\it Ginga}, due to its limited quality). We use the procedure described in Czerny et al.\ (2003) for calculating the SF and its uncertainty. The results are shown in Fig.\ \ref{f:sf}.

We clearly see the dips to the $\sim$150-d period and its subharmonics in the {\it RXTE}/ASM, BATSE and Ryle data. In the ASM data, the dip position is at $158\pm 14$ d. (Hereafter, the uncertainty has been estimated as equal to the FWHM of the absolute value of the dip profile after subtracting the surrounding continuum.) Thus, this result is consistent with our periodogram results. In the 20--100 keV BATSE data, the dip is at $153\pm 16$ d, while the 100--300 keV band shows only a shallow minimum at $\sim$150 d. The radio data show the dip at
$151\pm 19$ d. Interestingly, there is no sign of any dip at $\sim$192~d, which was found in the corresponding periodogram (see Section \ref{superorbital}). On the other hand, the \ariel show a pronounced dip at $\sim$276 d. The {\it Vela 5B\/} show only a very shallow minimum around $\sim$300 d.

Thus, the SF technique fully confirms the periodogram results for the {\it RXTE}/ASM and for the 20--100 keV BATSE band. Thus, we can consider those results as beyond any reasonable doubt. For other data, the SF results are in only partial agreement with the periodogram ones.

\bsp

\label{lastpage}

\end{document}

\setcounter{table}{1}
\setcounter{page}{16}

\begin{table*}

\caption{The results of the analysis of the orbital modulation in the hard state. See Section 3 for the definitions. The values of ${\cal P}$, $\Theta$, $n_{\rm corr}$ and $P_1$ are found directly from the periodograms whereas the remaining quantities are obtained for the folded and averaged lightcurves using the ephemeris of equation (10). See Table 3 for the values of $n$. The units of $\exp(G_0)$ are those shown in Fig.\ 1. }
\fontsize{7pt}{3.5mm} \selectfont

\label{t:orbital}
\begin{tabular}{lcrccrrrrrrrcr}
\hline
Detector &
${\cal P}$ [d] &
$\Theta$ &
$n_{\rm corr}$ &
$P_1$ &
$G_0$ &
$G_1$ &
$G_2$ &
$G_3$ &
$\varphi_1$ &
$\varphi_2$ &
$\varphi_3$ &
$G_{\rm max}$$-$$G_0$ &
$\phi_{\rm mod}$[\%]
\\
\hline
{\it RXTE}(1.5--3\,keV)        
                        & $5.5995\!\pm\! 0.0029$
                        & 171.1
			& 1.79
                        & $<10^{-10}$
			& $1.758\!\pm\! 0.004$
                        & $0.129\!\pm\! 0.007$
                        & $0.055\!\pm\! 0.006$
                        & $0.021\!\pm\! 0.006$
                        & $0.109\!\pm\! 0.046$
                        & $-0.177\!\pm\! 0.114$
                        & $-0.164\!\pm\! 0.299$
                        & 0.098
                        & $26.1\!\pm\! 1.2$
\\
{\it RXTE}(3--5\,keV)          
			& $5.5995\!\pm\! 0.0039$
                        & 108.7
			& 1.87
                        & $<10^{-10}$
			& $1.786\!\pm\! 0.003$
                        & $0.068\!\pm\! 0.005$
                        & $0.027\!\pm\! 0.005$
                        & $0.013\!\pm\! 0.005$
                        & $0.069\!\pm\! 0.073$
                        & $-0.331\!\pm\! 0.182$
                        & $0.050\!\pm\! 0.388$
                        & 0.058
                        & $15.3\!\pm\! 0.9$
\\
{\it RXTE}(5--12\,keV)         
			& $5.5994\!\pm\! 0.0050$
                        & 41.1
			& 1.90
                        & $<10^{-10}$
			& $2.165\!\pm\! 0.003$
                        & $0.035\!\pm\! 0.004$
                        & $0.012\!\pm\! 0.004$
                        & $0.008\!\pm\! 0.004$
                        & $-0.028\!\pm\! 0.107$
                        & $-0.473\!\pm\! 0.310$
                        & $-0.459\!\pm\! 0.461$
                        & 0.032
                        & $8.3\!\pm\! 1.0$
\\
Ryle(15 GHz)           
			& $5.6000\!\pm\! 0.0009$
                        & 148.3
			& 2.15
                        & $<10^{-10}$
			& $-4.456\!\pm\! 0.007$
                        & $0.154\!\pm\! 0.011$
                        & $0.038\!\pm\! 0.010$
                        & $-0.026\!\pm\! 0.010$
                        & $0.957\!\pm\! 0.063$
                        & $1.725\!\pm\! 0.261$
                        & $-0.039\!\pm\! 0.371$
                        & 0.144
                        & $30.4\!\pm\! 1.8 $
\\
GBI(8.30 GHz)
                        & $5.5997\!\pm\! 0.0019$
                        & 7.3
                        & 1.12
                        & $2\times 10^{-4}$
			& $-4.193\!\pm\! 0.010$
                        & $0.074\!\pm\! 0.014$
                        & --
                        & --
                        & $1.125\!\pm\! 0.202$
                        & --
                        & --
                        & 0.074
                        & $13.8\!\pm\! 2.4 $
\\
GBI(2.25 GHz)
                        & $5.5676\!\pm\! 0.0020$
                        & 4.0
                        & 1.19
                        & 0.02
			& $-4.255\!\pm\! 0.009$
                        & $0.023\!\pm\! 0.014$
                        & --
                        & --
                        & $1.986\!\pm\! 0.589$
                        & --
                        & --
                        & 0.023
                        & $4.5\!\pm\! 2.7 $
\\
{\it CGRO}(20--100\,keV)     
			& $5.5991\!\pm\! 0.0021$
                        & 3.7
			& 2.28
                        & 0.18
			& $2.319\!\pm\! 0.003$
                        & $0.016\!\pm\! 0.004$
                        & --
                        & --
                        & $-0.089\!\pm\! 0.235$
                        & --
                        & --
                        & 0.016
                        & $3.1\!\pm\! 0.8$
\\
{\it CGRO}(100--300\,keV)    
			& $5.5974\!\pm\! 0.0049$
                        & 1.0
			& 3.05
                        & 0.81
			& $1.836\!\pm\! 0.004$
                        & $0.010\!\pm\! 0.008$
                        & --
                        & --
                        & $-0.266\!\pm\! 0.506$
                        & --
                        & --
                        & 0.010
                        & $1.6\!\pm\! 1.6$
\\
{\it Ginga}(1--6\,keV)        
			& $5.6014\!\pm\! 0.0023$
                        & 12.6
			& 1.24
                        & $2\times 10^{-6}$
			& $-0.688\!\pm\! 0.016$
                        & $0.127\!\pm\! 0.024$
                        & --
                        & --
                        & $0.039\!\pm\! 0.173$
                        & --
                        & --
                        & 0.127
                        & $22.4\!\pm\! 3.7$
\\
{\it Ginga}(6--20\,keV)       
			& $5.6027\!\pm\! 0.0032$
                        & 4.9
			& 1.09
                        & $4\times 10^{-3}$
			& $-1.148\!\pm\! 0.018$
                        & $0.086\!\pm\! 0.027$
                        & --
                        & --
                        & $0.246\!\pm\! 0.292$
                        & --
                        & --
                        & 0.086
                        & $15.8\!\pm\! 4.5$
\\
\hline
\end{tabular}
\end{table*}

\setcounter{table}{2}

\begin{table*}
\caption{The most significant superorbital periodicities found in the hard state. $N=1$ in all cases, and $n$ gives the total number of points in the light curves. See Appendix A for discussion of the {\it Vela 5B} data. The values of ${\cal P}$, $\Theta$, $n_{\rm corr}$, $P_1$ and $P_{N_{\rm eff}}$ are found directly from the periodograms whereas the remaining quantities are obtained for the folded and averaged lightcurves with the ephemeris of equation (12) used for the first 11 rows. Since $N=1$ throughout, $G_{\rm max}-G_0 = G_1$. The units of $\exp(G_0)$ are those shown in Fig.\ 1.}
\label{t:super}
\begin{center}
\begin{tabular}{lcrrrcccrrrr}
\hline
Detector &
${\cal P}$ [d] &
$\Theta$ &
$n$ &
$n_{\rm corr}$ &
$P_1$ &
$N_{\rm eff}$ &
$P_{N_{\rm eff}}$ &
$G_0$ &
$G_1$ &
$\varphi_1$ &
$\phi_{\rm mod}$[\%] \\
\hline
\asmm(1.5--3 keV)         & $152.60\! \pm\! 1.20$
                          & 722.7
			  & 27232
                          & 1.80
                          & $<10^{-10}$
			  & --
                          & --
			  & $1.771\! \pm\! 0.015^1$
                          & $0.134\! \pm\! 0.021^1$
                          & $0.287\! \pm\! 0.169^1$
                          & $23.5\! \pm\! 3.2$\\
\asmm(3--5 keV)           & $153.02\! \pm\! 0.79$
                          & 1233.9
			  & 27344
                          & 1.89
                          & $<10^{-10}$
                          & --
                          & --
			  & $1.800\! \pm\! 0.012^1$
                          & $0.115\! \pm\! 0.016^1$
                          & $0.124\! \pm\! 0.148^1$
                          & $20.5\! \pm\! 2.5$\\
\asmm(5--12 keV)          & $153.10\! \pm\! 0.66$
                          & 1563.6
			  & 27345
                          & 1.91
                          & $<10^{-10}$
                          & --
                          & --
			  & $2.182\! \pm\! 0.009^1$
                          & $0.114\! \pm\! 0.013^1$
                          & $0.000\! \pm\! 0.122^1$
                          & $20.4\! \pm\! 2.1$ \\
Ryle(15 GHz)              & $150.69\! \pm\! 0.55$
                          & 257.30
			  & 12493
                          & 2.14
                          & $<10^{-10}$
                          & --
                          & --
			  & $-4.454\! \pm\! 0.008^1$
                          & $0.106\! \pm\! 0.012^1$
                          & $-0.041\! \pm\! 0.119^1$
                          & $19.1\! \pm\! 1.9$\\
GBI(8.30 GHz)             & $147.30\! \pm\! 1.03$
                          & 44.1
			  & 1547
                          & 1.12
                          & $<10^{-10}$
                          & --
                          & --
			  & $-4.184\! \pm\! 0.008^1$
                          & $0.124\! \pm\! 0.013^1$
                          & $0.325\! \pm\! 0.091^1$
                          & $22.0\! \pm\! 2.0$\\
GBI(2.25 GHz)              & $150.92\! \pm\! 0.59$
                          & 37.3
			  & 1547
                          & 1.18
                          & $<10^{-10}$
                          & --
                          & --
			  & $-4.247\! \pm\! 0.011^1$
                          & $0.107\! \pm\! 0.014^1$
                          & $0.329\! \pm\! 0.160^1$
                          & $19.3\! \pm\! 2.2$\\
{\it CGRO}(20--100 keV)
                          & $151.04\! \pm\! 0.32$
                          & 109.1
			  & 2511
                          & 2.28
                          & $<10^{-10}$
                          & --
                          & --
			  & $2.324\! \pm\! 0.005^1$
                          & $0.081\! \pm\! 0.007^1$
                          & $-0.003\! \pm\! 0.089^1$
                          & $14.9\! \pm\! 1.2$\\
{\it CGRO}(100--300 keV)
                          & $152.37\! \pm\! 0.51$
                          & 41.9
			  & 2511
                          & 3.05
                          & $<10^{-10}$
                          & --
                          & --
			  & $1.843\! \pm\! 0.007^1$
                          & $0.070\! \pm\! 0.008^1$
                          & $0.096\! \pm\! 0.141^1$
                          & $13.1\! \pm\! 1.4$ \\
{\it Ginga}(1--6 keV)
                          & $152.13\! \pm\! 1.55$
                          & 11.3
			  & 337
                          & 1.22
                          & $7\times 10^{-6}$
                          & --
                          & --
			  & $-0.717\! \pm\! 0.017^1$
                          & $0.109\! \pm\! 0.024^1$
                          & $-0.005\! \pm\! 0.223^1$
                          & $19.6\! \pm\! 3.9$ \\
{\it Ginga}(6--20 keV)
                          & $151.01\! \pm\! 1.52$
                          & 10.1
			  & 337 
                          & 1.19
                          & $2\times 10^{-5}$
                          & --
                          & --
			  & $-1.179\! \pm\! 0.015^1$
                          & $0.109\! \pm\! 0.024^1$
                          & $0.157\! \pm\! 0.208^1$
                          & $19.6\! \pm\! 3.9$ \\
{\it Ariel 5}(3--6 keV)
                          & $151.33\! \pm\! 0.79$
                          & 24.8
			  & 1972
                          & 1.30
                          & $<10^{-10}$
                          & --
                          & --
			  & $-1.053\! \pm\! 0.011^1$
                          & $0.074\! \pm\! 0.015^1$
                          & $0.477\! \pm\! 0.220^1$
                          & $13.8\! \pm\! 2.6$ \\
\hline
Ryle(15 GHz)              & $192.29\! \pm\! 1.04$
                          & 281.8
			  & 12493
                          & 2.12
                          & $<10^{-10}$
                          & 33.1
                          & $<10^{-10}$
			  & $-4.461\! \pm\! 0.011^2$
                          & $0.107\! \pm\! 0.011^2$
                          & $-0.406\! \pm\! 0.101^2$
                          & $19.3\! \pm\! 1.8$ \\
GBI(2.25 GHz)              & $185.47\! \pm\! 1.10$
                          & 27.9
			  & 1547
                          & 1.19
                          & $<10^{-10}$
                          & 17.8
                          & $<10^{-10}$
			  & $-4.257\! \pm\! 0.017^2$
                          & $0.052\! \pm\! 0.024^2$
                          & $0.559\! \pm\! 0.446^2$
                          & $9.9\! \pm\! 4.3$ \\
{\it Ginga}(1--6 keV)
                          & $230.30\! \pm\! 3.39$
                          & 17.3
			  & 337
                          & 1.26
                          & $2\times 10^{-8}$
                          & 21.4
                          & $4\times 10^{-7}$
			  & $-0.701\! \pm\! 0.018^3$
                          & $0.106\! \pm\! 0.026^3$
                          & $1.191\! \pm\! 0.236^3$
                          & $19.1\! \pm\! 4.2$ \\
{\it Ginga}(6--20 keV)
                          & $232.65\! \pm\! 3.35$
                          & 11.8
			  & 337
                          & 1.11
                          & $1\times 10^{-6}$
                          & 20.9
                          & $2\times 10^{-5}$
			  & $-1.139\! \pm\! 0.014^3$
                          & $0.042\! \pm\! 0.019^3$
                          & $0.615\! \pm\! 0.517^3$
                          & $8.1\! \pm\! 3.5$ \\
{\it Ariel 5}(3--6 keV)
                          & $278.55\! \pm\! 3.07$
                          & 81.7
			  & 1972
                          & 1.31
                          & $<10^{-10}$
                          & 18.6
                          & $<10^{-10}$
			  & $-1.074\! \pm\! 0.011^3$
                          & $0.135\! \pm\! 0.015^3$
                          & $1.407\! \pm\! 0.131^3$
                          & $23.7\! \pm\! 2.3$ \\
{\it Vela 5B}(3--12 keV)
                          & $289.50\! \pm\! 1.70$
                          & 19.3
			  & 1405
                          & 1.20
                          & $<10^{-10}$
                          & 42.8
                          & $1\times 10^{-8}$
			  & $4.721\! \pm\! 0.003^3$
                          & $0.023\! \pm\! 0.004^3$
                          & $1.164\!\pm\! 0.160^3$
                          & $4.5\!\pm\! 0.7$ \\
\hline
\end{tabular}
\end{center}
\begin{flushleft}
\vspace*{-5pt}
\hspace*{80pt}
$^1T_0 = \mbox{MJD} \ 50514.59$, $G_1$ and $\varphi_1$ given for the common 
period of 151.43~d.
\\
\hspace*{80pt}
$^2T_0 = \mbox{MJD} \ 50514.59.$
\\
\hspace*{80pt}
$^3T_0 = \mbox{MJD} \ 44450.00.$
\end{flushleft}
\end{table*}